\documentclass{article} % For LaTeX2e
\usepackage{iclr2027_arxiv,times}

\usepackage{amsmath,amsfonts,bm}

\def\eqref#1{equation~\ref{#1}}
\def\1{\bm{1}}

\DeclareMathAlphabet{\mathsfit}{\encodingdefault}{\sfdefault}{m}{sl}
\SetMathAlphabet{\mathsfit}{bold}{\encodingdefault}{\sfdefault}{bx}{n}

\usepackage{hyperref}
\usepackage{url}
\usepackage{booktabs} 
\usepackage{subcaption}
\usepackage{graphicx}
\usepackage{adjustbox}
\usepackage{array}
\usepackage{tabularx}
\usepackage{svg}

\usepackage{xcolor}
\usepackage{listings}
\usepackage[most]{tcolorbox}
\tcbuselibrary{breakable,skins,listings}

\definecolor{appendixboxframe}{RGB}{188,188,188}
\definecolor{appendixboxtitle}{RGB}{188,188,188}
\definecolor{appendixboxback}{RGB}{252,252,252}

\definecolor{codeframe}{RGB}{190,190,190}
\definecolor{codetitle}{RGB}{190,190,190}
\definecolor{codeback}{RGB}{253,253,253}

\newtcolorbox{promptbox}[1]{
    enhanced,
    breakable,
    colback=appendixboxback,
    colframe=appendixboxframe,
    colbacktitle=appendixboxtitle,
    coltitle=white,
    title={#1},
    fonttitle=\bfseries,
    fontupper=\footnotesize,
    boxrule=0.8pt,
    arc=3pt,
    outer arc=3pt,
    left=8pt,
    right=8pt,
    top=7pt,
    bottom=7pt,
    before skip=8pt,
    after skip=10pt
}

\newtcolorbox{sourceexample}[1]{
    enhanced,
    breakable,
    colback=gray!3,
    colframe=gray!40,
    colbacktitle=gray!35,
    coltitle=white,
    title={#1},
    fonttitle=\bfseries,
    fontupper=\footnotesize,
    boxrule=0.7pt,
    arc=3pt,
    outer arc=3pt,
    left=8pt,
    right=8pt,
    top=6pt,
    bottom=6pt,
    before skip=7pt,
    after skip=9pt
}

\newtcolorbox{strategybox}[1]{
    enhanced,
    breakable,
    colback=appendixboxback,
    colframe=appendixboxframe,
    colbacktitle=appendixboxtitle,
    coltitle=white,
    title={#1},
    fonttitle=\bfseries,
    fontupper=\small,
    boxrule=0.8pt,
    arc=3pt,
    outer arc=3pt,
    left=8pt,
    right=8pt,
    top=7pt,
    bottom=7pt,
    before skip=8pt,
    after skip=10pt
}

\newtcblisting{codebox}[1]{
    enhanced,
    breakable,
    listing only,
    listing engine=listings,
    listing options={style=academicpython},

    colback=codeback,
    colframe=codeframe,
    boxrule=0.65pt,
    arc=2pt,
    outer arc=2pt,

    colbacktitle=codetitle,
    coltitle=white,
    title={#1},
    fonttitle=\bfseries\footnotesize,

    left=5pt,
    right=5pt,
    top=3pt,
    bottom=3pt,
    boxsep=2pt,

    before skip=6pt,
    after skip=7pt,

    toptitle=2.5pt,
    bottomtitle=2.5pt
}

\newtcolorbox{policybox}[1]{
    enhanced,
    breakable,
    width=0.92\linewidth,
    center,
    colback=white,
    colframe=gray!45,
    colbacktitle=gray!42,
    coltitle=white,
    title={#1},
    fonttitle=\small\bfseries,
    fontupper=\footnotesize,
    boxrule=0.7pt,
    arc=3pt,
    outer arc=3pt,
    left=10pt,
    right=10pt,
    top=7pt,
    bottom=7pt,
    before skip=8pt,
    after skip=10pt
}

\title{AlphaOpsBench: Benchmarking End-to-End Alpha Strategy Operationalization in Prediction Markets}

\author{
\textbf{Huaiyu Jia}$^{1}$ \quad
\textbf{Mingxuan Zhao}$^{1}$ \quad
\textbf{Jincheng Gao}$^{1}$ \quad
\textbf{Zifan Peng}$^{1}$ \quad
\textbf{Wentao Zhang}$^{2}$ \quad
\textbf{Siguang Li}$^{1}$ \quad
\textbf{Shuo Sun}$^{1}$ \\[4pt]
$^{1}$The Hong Kong University of Science and Technology (Guangzhou), Guangzhou, China \\
$^{2}$Nanyang Technological University, Singapore
}

\iclrfinalcopy % Uncomment for camera-ready version, but NOT for submission.
\begin{document}

\maketitle

\begin{abstract}
Large language models increasingly generate quantitative trading strategies, yet existing benchmarks assume standardized assets, numerical features, or directly compilable strategy representations---assumptions that prediction-market strategies violate, since a coarse idea may leave the traded outcome, causal information source, signal definition, threshold, sizing, order policy, exit, and settlement behavior unspecified. We introduce \textsc{AlphaOpsBench}, which evaluates end-to-end operationalization from source-grounded economic hypotheses to auditable executable programs over 581 source-preserving strategy records and a lifecycle-scale Polymarket dataset with 1.28 million binary markets, 183.6 million cleaned executions, settlement evidence, and limit-order-book history, comparing Direct generation against a Staged design-then-code protocol. In a corrected independent-generation study over 36 controlled tasks and 24 preregistered real strategies, strict end-to-end validity remains rare: Direct and Staged obtain 35/180 and 20/180 canonical passes on the controlled cohort and no confirmed pass on the real cohort, and repeated generations vary substantially in model-owned economic choices. By contrast, 775,725 of 783,655 scheduled historical replays complete, showing that replayability is a far weaker property than source-faithful operationalization. Financial outcomes depend on the declared execution model and available historical evidence, and fee and liquidity experiments show that execution costs alter subsequent trading paths rather than acting only as ex-post deductions. \textsc{AlphaOpsBench} thus separates strategy fidelity, behavioral validity, historical executability, and financial performance in an evidence-aware benchmark for LLM-based quantitative research in prediction markets.
\end{abstract}

\section{introduction}
Prediction markets provide a market-based mechanism for aggregating dispersed information about uncertain future events \citep{ng2026price,jia2026unlocking}. By enabling participants to trade contracts whose payoffs are contingent on event outcomes, market prices distill collective beliefs and serve as continuously updated probabilistic forecasts. This mechanism has been extensively studied across domains including elections, macroeconomic indicators, sports, and other quantitatively traded settings \citep{reichenbach2025exploring}. Crucially, prediction markets tightly couple prediction with financial decision-making: participants must not only estimate what is likely to occur but also assess whether the discrepancy between their beliefs and prevailing market prices is sufficiently large to warrant a trade \citep{kalshi2026, predictfun2026}. Modern blockchain-based prediction markets further offer programmable access to event-linked contracts, well-defined resolution structures, observable market dynamics, and verifiable settlement, rendering them a natural testbed for studying information aggregation, quantitative trading, and autonomous financial agents (e.g., Polymarket \citep{polymarket2026}, Robinhood \citep{robinhoodpredictionmarkets2026}). Prediction markets thus constitute a unique research environment in which beliefs about real-world events are continuously translated into prices, positions, and ultimately realized returns \citep{yu2026building}.

Despite the importance of prediction market strategies, their systematic generation and evaluation remain underexplored. Recent work leveraging LLMs has made significant progress in formulaic alpha mining, strategy generation, and automated backtesting; however, these approaches typically assume a predefined asset universe, data schema, or numerical strategy representation. Prediction market strategies are inherently unstandardized. Strategies such as buying underpriced outcomes, following informed traders, or exploiting inconsistencies across related contracts express economic hypotheses rather than directly compilable rules: they often leave underspecified the relevant events and outcomes, probability estimates, information timestamps, decision thresholds, position sizing, order placement policies, and exit conditions. Consequently, such strategies generally cannot be reduced to equity-style expressions such as momentum or mean-reversion formulas. Generating code directly from these descriptions introduces additional challenges, as the model must simultaneously operationalize the underlying economic rationale, reconcile event-specific and point-in-time data, preserve the semantics of contracts and outcomes, and implement effective trading behavior. As a result, a generated program may be executable yet encode a strategy that is semantically divergent or causally invalid. To the best of our knowledge, no existing benchmark specifically evaluates the end-to-end translation from coarse-grained prediction market strategies to faithful, executable code, leaving a notable gap in benchmarking for prediction market strategy design and code generation.

\begin{figure}[t]
    \centering
    \includegraphics[width=\linewidth]{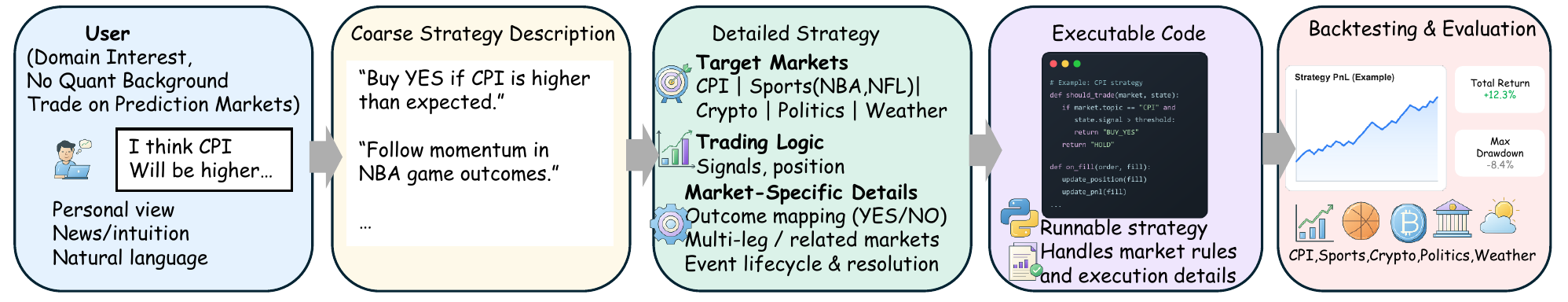}
    \caption{Motivation: converting a coarse natural-language trading idea into an executable prediction-market program requires progressively fixing target markets, signals, position rules, outcome semantics, and lifecycle constraints---AlphaOpsBench.}
    \label{fig:strategy_operationalization_motivation}
\end{figure}

To address these limitations, we introduce \textbf{AlphaOpsBench}, a benchmark that shifts evaluation from isolated strategy description and black-box financial scoring to end-to-end operationalization: can an LLM turn a real-world prediction-market trading idea into an executable, testable program? Prediction markets resist the homogeneous-asset assumption of conventional financial benchmarks---each event contract carries its own outcomes, information process, lifecycle, and settlement---so we build the benchmark at lifecycle scale, combining strategy descriptions with market metadata, execution records, settlement evidence, and historical microstructure. AlphaOpsBench evaluates the full chain from strategy understanding and decision formalization to executable policy generation in a unified environment, testing whether LLMs can act as quantitative strategy designers rather than producers of plausible trading descriptions.

We further connect generated strategies to realistic historical execution through two complementary backtesting frameworks: a broad replay engine that evaluates behavior across a large market universe, and an execution-aware engine that incorporates order-book state, transaction dynamics, and market-specific constraints. This pairing separates strategy-generation capability from execution feasibility and financial outcome. Throughout, the framework preserves the full decision chain---source intent, model-owned operational choices, executable code, historical execution and evaluation, providing a reproducible foundation for end-to-end pipeline in prediction markets.

\section{Related Work}
\label{sec:related_work}

\subsection{Financial Knowledge and Trading Benchmarks}
\label{sec:related_financial_benchmarks}

Financial benchmarks have evolved from static document reasoning to autonomous quantitative research and executable trading. Early works established numerical and hybrid reasoning over financial reports via FinQA~\citep{chen2021finqa}, TAT-QA~\citep{zhu2021tat}, and ConvFinQA~\citep{chen2022convfinqa}. Subsequent studies expanded to open-domain retrieval~\citep{hu2026finsearchcomp}, multi-tool crypto analysis~\citep{eswaran2026cryptoanalystbench}, time-series reasoning~\citep{gwiazda2026timeseriesexamagent}, and test-time forecasting~\citep{kayaalp2026test}. As LLMs transition into active quantitative researchers, systems have been developed for arbitrage~\citep{roy2026alphasearch}, high-frequency deployment~\citep{song2026timi}, and automated alpha discovery through skill accumulation~\citep{wang2026factorminer}, feedback optimization~\citep{wang2026alphamaster}, agentic evolution~\citep{tang2026alphaagentevo}, and GFlowNets~\citep{chen2026alphasage}. To evaluate these pipelines, recent benchmarks standardized formulaic alpha mining~\citep{luo2026alphabench}, multi-dimensional factor assessment~\citep{ding2026alphaeval}, executable strategy generation~\citep{zhang2026alphaforgebench}, and automated backtesting~\citep{wang2026backtestbench}. However, these evaluations typically assume hypotheses can be reduced to compact numerical relations, neglecting the operationalization of underspecified economic decisions, temporal alignment, and execution semantics. A benchmark must therefore assess whether generated programs are semantically faithful to the original strategy and valid under real-world market constraints, rather than merely executable.

\subsection{Prediction Markets and Blockchain-based Trading}
While existing benchmarks evaluate financial search and multi-tool crypto analysis~\citep{hu2026finsearchcomp}, or trading agents via deterministic historical replay~\citep{arora2026predictionmarketbench}, none address end-to-end operationalization of incomplete prediction-market strategies. Complementary work on on-chain AMM forecasting~\citep{jia2026towards}, micro-level transaction understanding~\citep{peng2025txsum}, and cross-chain interoperability~\citep{cao2026price} similarly overlooks this gap. On Polymarket, contingent claims require operations such as collateral splitting, position merging, and redemption under market-specific rules~\citep{arora2026predictionmarketbench,ng2026price}. Since such strategies couple event-specific evidence, contract semantics, and lifecycle actions, direct code generation can silently alter the intended economic hypothesis; we therefore benchmark their translation into auditable, executable prediction-market strategies.

\section{AlphaOps in the Prediction Market}

\subsection{Dataset Construction}
\textbf{Data Engine in the Polymarket.} To operationalize the quantification pipeline and evaluate factor performance in prediction markets, we focus on Polymarket \citep{polymarket2026} as our experimental platform, abstracting away from complex scenarios such as cross-market and cross-exchange arbitrage. Polymarket constitutes an ideal subject for this study, owing to its mature mechanism design and the comprehensive public availability of on-chain transaction data \citep{polymarketGammaAPI2026,polygonRPC2026}. To establish a comprehensive foundation for quantitative research in prediction markets, we construct a lifecycle-scale dataset by integrating publicly accessible market metadata, on-chain execution and settlement records (including oracle resolutions), and continuously collected limit-order-book (LOB) feeds. We specifically focus on the three-month period from June 1 to September 1, 2026. This window is selected because Polymarket's infrastructure and trading protocols had reached a mature and stable operational regime, effectively mitigating the strategy invalidation risks associated with earlier chain-level vulnerabilities or frequent protocol iterations \citep{shen2026ghosts}. Over this period, we aggregated extensive raw transaction and order data. Following rigorous standardization, cross-source reconciliation, and quality filtering, the finalized dataset comprises 1,276,684 binary markets and 183,625,793 execution records, with its corresponding settlement Oracle data and LOB data. Detailed procedures regarding data acquisition, normalization, and quality control are provided in Appendix~\ref{data_construction}.

\textbf{Real-world Strategy Collection.} We collect 581 raw strategy records from five complementary sources: a hand-curated strategy guide (128), ChatGPT (119), Qwen (120), Grok (112), and academic literature (102 records from 35 papers on prediction markets and forecasting). Each record is stored as an immutable, hash-bound \textsc{SourceRecord} that separates explicit requirements---data inputs, decision rules, trading actions, and exit/settlement conditions---from open decisions, so underspecification is measured rather than silently repaired. Mechanism-level deduplication is applied to the 479 catalog records, merging entries with the same payoff source, primary signal, and entry--exit state model while preserving distinct recipes when data contracts, state models, position construction, or execution economics differ; this yields 106 canonical mechanism families and 27 shared components, with full lineage. The 102 literature records are deduplicated separately and merged into the same source-preserving pipeline. Retained records are compiled into 511 candidate \textsc{StrategyTask}s under explicit intake states; among 477 active task identities, 430 become frozen task contracts and 47 are skipped with logged blocking reasons. We grade each frozen task by specification completeness: an \emph{L1} source fixes the economic mechanism and all nine execution-relevant decisions---data contract, factor definition, window, threshold, direction, order strategy, position sizing, exit rule, and no-trade rule---so the task reduces to faithful translation; an \emph{L2} source states the mechanism and decision rules but leaves parameters such as thresholds, windows, and position sizes open within stated bounds; an \emph{L3} source provides only a high-level objective and an admissible design space, requiring end-to-end design. For controlled evaluation, we construct 12 diagnostic tasks from four mechanism cards---complete-set arbitrage, directional price continuation, inventory-constrained quoting, and trade-flow following---each anchored to a real task and instantiated at all three levels, thereby isolating specification completeness from strategy mechanism.

\subsection{End-to-End Pipeline}
Our evaluation converts written trading ideas into executable prediction-market strategies through five stages. Conventional pipelines map OHLCV bars to BUY/SELL/HOLD and abstract execution away; prediction markets instead make the trade lifecycle part of the strategy: YES/NO tokens split from and merge into collateral, orders carry time-in-force and partial-fill semantics, multi-leg strategies lack guaranteed joint execution, and payoff arrives only after oracle resolution and redemption. A generated strategy must therefore specify the traded outcome token, how complementary claims are combined, per-leg pricing and sizing, and settlement timing. 

\textbf{Step 1: Coarse strategy extraction.} Descriptions from guides, catalogs, and literature become coarse tasks preserving source references and explicit requirements across data inputs, mathematical relationships, trading actions, and market structure. Prediction-market strategies often prescribe action relationships that standard pipelines discard. 

% Our binary complete-set arbitrage task specifies buying equal YES and NO quantities when their combined ask plus costs falls below one collateral unit, then merging or holding to settlement. Extraction must retain both legs, equal sizing, fill-or-kill conditions, and alternative exits. The related split-and-sell strategy creates complementary claims before selling them—same payoff, different sequence. We therefore preserve outcome identities, action ordering, and required-versus-alternative distinctions.

\textbf{Step 2: Prompt construction.} Each task enters a standardized prompt with its description, requirements, data definitions (meaning, units, windows, scope), and trading capabilities (outcome selection, orders, time-in-force, protocol operations). Unlike price-series interfaces, prediction markets distinguish trade observations from order-book state. Our microprice quotation task requires bid/ask prices with depths to locate the quotation center; transaction volume alone provides neither. An observed YES trade does not establish the executable NO ask or its quantity. 

\textbf{Step 3: Detailed strategy design.} The model refines coarse tasks into explicit policies naming target markets, outcomes, observations, features, trading conditions, order parameters, and entry/exit/no-trade rules. Outcome identity is an economic decision, not a neutral parameter. Our extreme-probability strategy permits selling YES above 0.98 or buying NO below 0.02—alternatives requiring different price inputs, order directions, and inventory conditions. A design must select one route and specify its order logic and subsequent exits. Strategies buying entire outcome baskets must identify the mutually exclusive, exhaustive outcome set. Staged returns design without code; the design is validated against supported inputs and actions, then mapped to a common representation preserving economic choices.

\textbf{Step 4: Code generation and validation.} Staged implements its validated design in a second call; Direct supplies implementation with design. Both produce programs mapping decision-time observations and account state (holdings, outstanding orders) to outcome-specific actions. Validation separates source adherence from design–implementation agreement: code may faithfully implement an altered strategy. In the complete-set example, dropping the NO leg or using bid prices for acquisition costs changes the strategy even if the program executes. Static checks enforce interface and data-access contracts; behavioral tests probe outcome bindings, numerical boundaries, action sequences, missing inputs, and position-dependent behavior against independently derived expectations. Continuous cases examine subsequent decisions using account updates from the execution simulator, including exits after partial fills.

\textbf{Step 5: Backtest-based assessment.} Conventional OHLCV backtests abstract away market impact---ignoring slippage or assuming infinite depth---assumptions incompatible with prediction markets' thin liquidity and complex execution semantics (IOC, GTC, FOK). We therefore evaluate executable candidates only on active, liquid historical markets selected for applicability, activity, and data availability, sharing market data, capital, fees, and execution settings while maintaining independent accounts, using two bespoke engines: \textsc{Fill-Only V3}, a trade-tick simulator that models finite executable liquidity from on-chain fills under synthetic-arrival assumptions, and \textsc{PML2}, a limit-order-book replay engine that reconstructs depth and simulates marketable and resting orders with partial fills, cancellations, and expiry. Execution-level assessment is essential here: a combined YES--NO ask below one collateral unit does not guarantee both legs can be acquired at that price, and per-order fill-or-kill does not ensure joint execution. Both engines distinguish submitted orders from simulated fills, separate decision-time information from execution evidence, and isolate terminal valuation from settlement proceeds; zero-fill runs enter coverage statistics rather than return comparisons. Details are provided in Appendix~\ref{backtest}.

\subsection{Evaluation Methodology}
\label{sec:evaluation_pipeline}
We evaluate generated strategies along four dimensions matching the stages of operationalization. (1) Generation pipeline: coverage, design formation, code availability, preservation of source-grounded decisions, and behavioral correctness, with failures attributed separately to strategy interpretation, implementation constraints, or missing execution evidence, since a syntactically valid program need not preserve the intended economic mechanism. (2) Prediction-market reasoning: structured diagnostic tasks at three specification-completeness levels test outcome semantics, causal information timing, position transitions, and settlement constraints beyond general financial programming. (3) Historical execution: two complementary replay settings---a broad market-universe framework and an execution-aware framework---evaluate candidates financially only when sufficient execution evidence exists; incomplete inputs, unobservable executions, and invalid conditions are reported as distinct states rather than zero performance, and metrics cover returns, risk exposure, execution characteristics, and cost sensitivity to separate strategy quality from execution effects. (4) Stability: repeated generation measures consistency in strategy structure, behavioral validation, and downstream outcomes. All statistics carry valid sample sizes and failure categories, and results are organized by completeness level, task and market domain, and historical trajectories, distinguishing observed success from unresolved cases and unavailable evidence.

\begin{figure}
    \centering
    \includegraphics[width=1\linewidth]{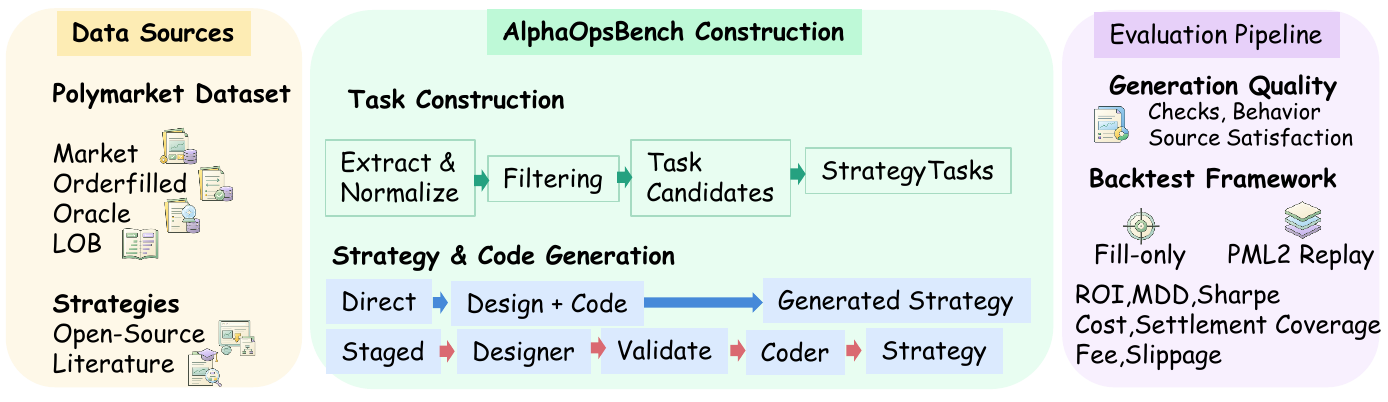}
    \caption{pipeline}
    \label{fig:pipeline}
\end{figure}

\section{Experiments}

\subsection{Experimental Settings}
\label{sec:experimental_settings}

\paragraph{Experimental design.}
The source-derived benchmark contains 477 active strategy tasks, which define
the main coverage denominator. We compare \textbf{Direct} and \textbf{Staged}
generation with Qwen3.8-27B at temperature $T=0.7$, using one generation per
task and method ($R=1$, $K=1$). This gives 954 planned task--method slots;
generation is attempted for 476 tasks, while one task is retained as an
intake exclusion. Direct produces the strategy specification and implementation
jointly, whereas Staged first produces an explicit specification and then
generates code from that design. Both methods use the same task descriptions,
source requirements, data interfaces, and trading capabilities.

Supplementary diagnostics are evaluated separately from the 477-task main
cohort. A 12-task suite spanning four mechanisms and three specification
levels (L1/L2/L3) is evaluated at $T\in\{0,0.7\}$ with five code samples per
condition, yielding 240 samples under the legacy protocol. These results are
used only as controlled diagnostics, since the Staged samples share one design
within each task--temperature condition and the legacy protocol has documented
prompt and evaluator limitations. Independent full-chain $R=5$ experiments
form a separate stability track and are not included in the main-cohort counts.

\paragraph{Backtest configuration.}
Historical replay covers the three-month period from June 1 to September 1,
2026, using Polymarket trades, available L2 order-book histories,
market--token mappings, and lifecycle and resolution records. Replay is
conditional on the availability of executable research artifacts rather than
on membership in the full 477-task corpus. The PML2 cohort contains 456
method-specific candidate packages covering 390 distinct tasks, while the
supported Fill-only V3 cohort contains 334 packages covering 324 tasks.
PML2 models order-book depth, latency, and order-lifecycle constraints,
whereas V3 provides broader trade-driven replay under a controlled
executable-liquidity assumption.

Each candidate--market--window replay uses an independent account initialized
with 1,000 USDC and no outcome-token inventory, with decisions made every
60 seconds. V3 windows extend to at most 48 hours, while the standard PML2
setting uses two-hour windows. Financial evaluation is further restricted to
episodes with valid execution and valuation evidence; replay coverage and
financial eligibility are therefore reported with separate denominators.

\paragraph{Execution sensitivity.}
We evaluate transaction costs and execution constraints in a separate paired
study that holds the candidate and historical window fixed while varying
fees, order size, entry latency, executable-liquidity capacity, and initial
capital. This study contains 529 replay configurations across 19
task--market windows from 15 tasks. We additionally perform a fee sweep over
213,263 previously evaluated episodes using fixed execution paths. The paired
replays capture changes in subsequent execution caused by fees or capital
constraints, whereas the population-scale sweep isolates the accounting
effect of transaction costs.

\paragraph{Evaluation metrics.}
We report generation, execution, and financial outcomes separately.
Generation metrics include response and design coverage, code availability,
static and behavioral validation, source-requirement checks, and final
candidate acceptance, each with its corresponding denominator. Execution
metrics include replay completion, input availability, order submission,
simulated fills, economic exposure, valuation availability, and settlement
status. For financially evaluable episodes, we report PnL, return on initial
capital (\textbf{ROI}), maximum drawdown (\textbf{MDD}), positive-return
incidence, and trade and fill statistics.

\subsection{Operationalization Accuracy and Stability}
\label{sec:operationalization_results}

\paragraph{Strict end-to-end validity remains challenging.}
The corrected independent-generation experiment contains 600 full-chain
attempts: 360 over controlled tasks and 240 over preregistered real-world
strategies. On the controlled cohort, Direct achieves 35 canonical passes
among 180 attempts and Staged achieves 20/180. At the task level, at least one
of five independent generations passes for 13/36 Direct tasks and 10/36
Staged tasks, while all five repetitions pass for only 2/36 and 1/36 tasks,
respectively. None of the 24 real-world strategies obtains a confirmed
canonical pass under either method. These results show that producing a
research-executable artifact is substantially easier than satisfying the full
source-fidelity, implementation, and behavioral contract.

Specification level alone does not induce a monotone ordering. Neither method
obtains a canonical pass on L1, whereas Direct obtains 17/60 and 18/60 passes
on L2 and L3 and Staged obtains 9/60 and 11/60. This does not imply that the
less specified tasks are easier: the levels expose different evidence
requirements and different proportions of source-given versus model-owned
economic decisions. We therefore report preservation and completion together
with their verification coverage rather than treating unverified decisions as
failures.

\paragraph{Repeated generations diverge primarily in economic completion.}
Independent repetitions also reveal that strategy stability is not well
characterized by code identity alone. We compare nine normalized economic
design slots---data, factor, window, threshold, direction, order policy,
sizing, exit, and no-trade behavior---within each task. Higher-level choices
such as trade direction and abstention are generally more reproducible than
thresholds, sizing, and several data-dependent choices, although agreement
must be interpreted jointly with the number of observable repetition pairs.

The same distinction appears at the behavioral level. On candidate-independent
common states, task-equal exact-action agreement for Direct/Staged is
\(95.0\%/90.3\%\) at L1, \(52.6\%/63.9\%\) at L2, and
\(67.5\%/36.9\%\) at L3. Restricting evaluation to states in which the
reference behavior is active yields the same qualitative conclusion: repeated
operationalizations can agree on broad strategy intent while diverging on the
precise conditions and actions that create exposure. Because jointly
observable outputs are not available for every planned pair, all agreement
statistics are reported together with pair coverage.

\begin{figure}
    \centering
    \includegraphics[width=1\linewidth]{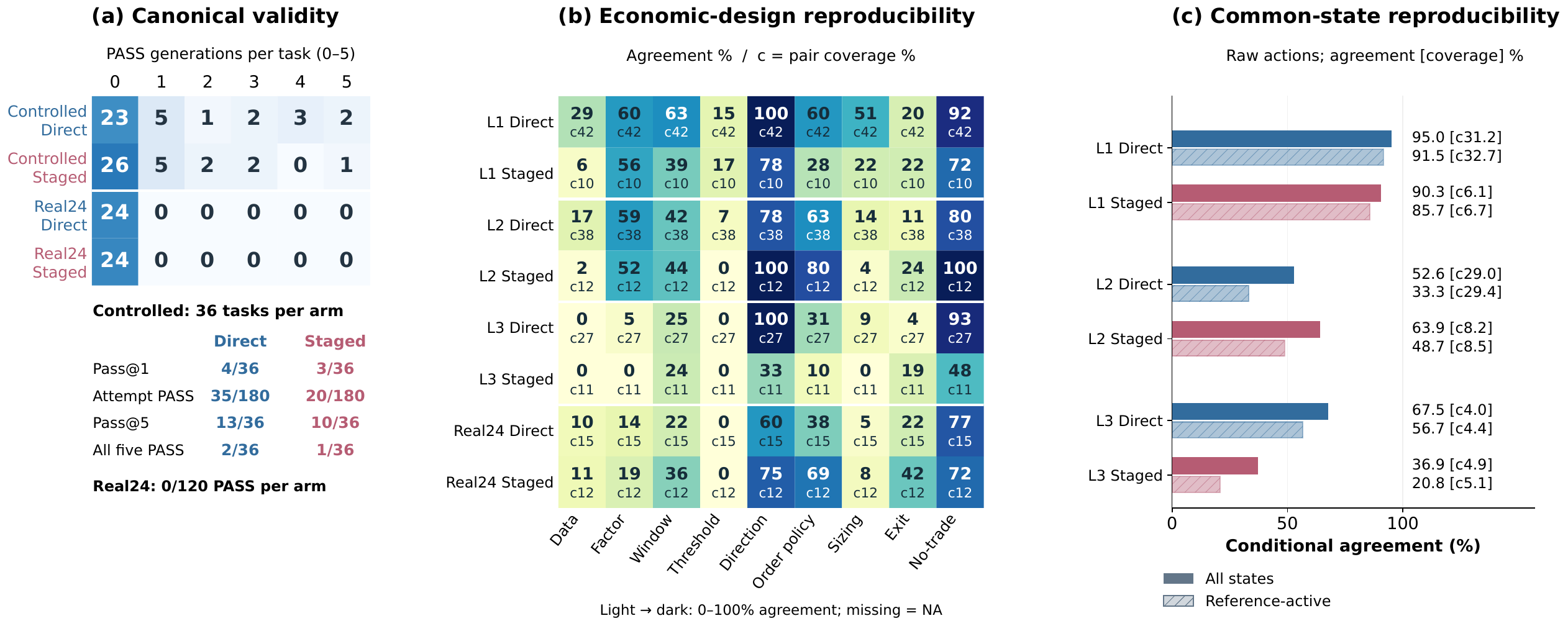}
    \caption{
\textbf{Operationalization validity and reproducibility under independent full-chain generation.}
\textbf{(a)} Canonical validity across five independent generations (\(R=5, K=1, T=0.7\)) for the 36 controlled tasks and 24 preregistered real-world tasks, showing the number of passing implementations obtained per task and the corresponding Pass@1, attempt-level success, Pass@5, and all-five success rates.
\textbf{(b)} Reproducibility of nine normalized economic design decisions---data, factor, window, threshold, direction, order policy, sizing, exit, and no-trade behavior. Cell values report conditional pairwise agreement, with jointly observable pair coverage shown separately in each cell.
\textbf{(c)} Common-state behavioral reproducibility for raw candidate actions on candidate-independent market and account states, reported over all evaluable states and over reference-active states. Agreement is task-equal and is always interpreted jointly with pair coverage; missing or unevaluable outputs are not treated as disagreements.
}
\label{fig:operationalization_stability}
\end{figure}

\subsection{Historical Execution and Financial Outcomes}
\label{sec:historical_results}

\paragraph{Replay coverage.}
We next evaluate the subset of generated artifacts for which an applicable
historical research-replay route is available. Across PML2 and V3, the
experiment schedules 783,655 replay units and obtains 775,725 COMPLETE
executions. PML2 completes all 377,255 Direct replays and 74,998 of 75,000
Staged replays. V3 completes 313,657 of 321,400 Direct replays and 9,815 of
10,000 Staged replays. Complete execution does not imply financial
evaluability: zero-fill episodes and executions lacking sufficient terminal
valuation remain in the coverage denominator but are excluded from return
aggregation.

\paragraph{Financial outcomes.}
Under task-equal aggregation, the mean window-end Net ROI is
\(-3.476\%\) for PML2--Direct and \(-3.440\%\) for PML2--Staged.
The corresponding values under the V3 research setting are
\(+0.836\%\) and \(+0.646\%\). The task-block 95\% intervals are
\([-3.515,-3.436]\%\), \([-3.504,-3.386]\%\),
\([0.814,0.857]\%\), and \([0.496,0.788]\%\), respectively.
These results characterize different routed replay populations and execution
contracts.

\begin{figure}
    \centering
    \includegraphics[width=1\linewidth]{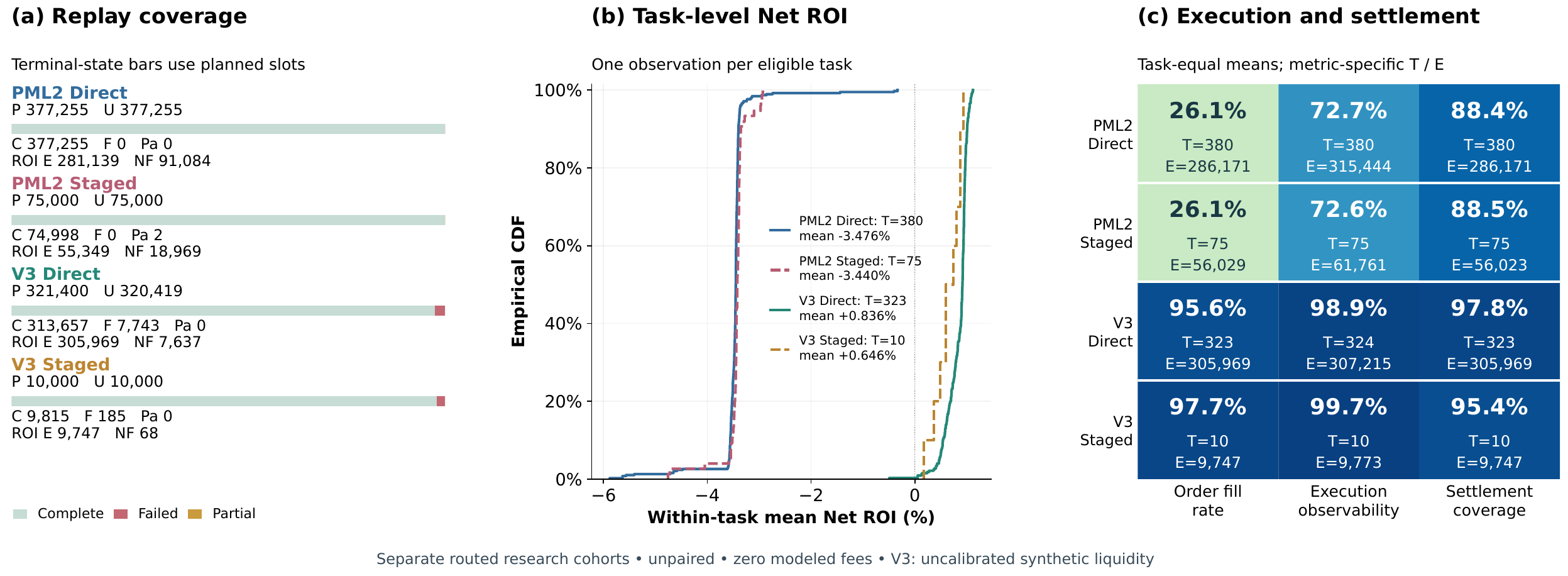}
    \caption{
\textbf{Historical replay coverage, financial outcomes, and execution evidence for research-executable strategies.}
\textbf{(a)} Replay coverage for the four routed engine--method cohorts, reporting planned and unique execution counts together with COMPLETE, FAILED, and PARTIAL terminal states; financially evaluable and no-fill counts are shown separately because they are not mutually exclusive engine states.
\textbf{(b)} Empirical distributions of within-task mean Net ROI, with one observation per financially evaluable task rather than per replay episode.
\textbf{(c)} Task-equal order-fill rate, execution observability, and settlement coverage, with metric-specific task and execution denominators.
PML2 and V3 operate on different routed candidate populations and execution contracts; the panels therefore characterize each research replay setting and should not be interpreted as a paired cross-engine performance comparison. V3 uses uncalibrated synthetic executable liquidity in this experiment.
}
\label{fig:historical_execution}
\end{figure}

\paragraph{Risk, execution, and settlement evidence.}
Window-end return alone does not characterize prediction-market execution.
For PML2, task-equal MDD is \(3.520\%\) for Direct and \(3.481\%\) for
Staged, with settlement coverage of \(88.35\%\) and \(88.50\%\).
Their order-fill rates are approximately \(26\%\), reflecting the stronger
order-book and readiness constraints of the PML2 replay. V3 exhibits much
higher simulated order-fill rates (\(95.57\%\) and \(97.67\%\)) and execution
observability (\(98.91\%\) and \(99.74\%\)), but these quantities arise under
a different modeled-liquidity contract and should not be interpreted as
evidence that V3 provides superior execution.

Metric availability is itself heterogeneous. For example, Net ROI is
recoverable for more than \(95\%\) of the main V3 replay population, whereas
quantity-fill evidence is available for only about one quarter of the same
planned units. Main-cohort input-coverage and markout statistics cannot be
reconstructed from the archived evidence and are therefore reported as
unavailable rather than zero. We provide the complete metric-specific
availability matrix in the appendix.

\subsection{Transaction fees and Slippage in the Polymarket}

\begin{table*}[t]
\centering
\scriptsize
\setlength{\tabcolsep}{2.6pt}
\renewcommand{\arraystretch}{1.08}

% ============================================================
% (a) Cash-feedback
% ============================================================
\begin{minipage}[t]{0.32\textwidth}
\vspace{0pt}
\centering
\textbf{(a) Fee-induced cash feedback}
\vspace{3pt}

\begin{adjustbox}{max width=\linewidth}
\begin{tabular}{llrrrr}
\toprule
\textbf{Engine} & \textbf{Fee} & \textbf{Capital}
& \textbf{Changed/N} & \textbf{Feedback} & \textbf{Max gap} \\
\midrule
PML2 & \(r=.07\) & 20   & 1/5 & +0.834 & 4.168 \\
PML2 & \(r=.07\) & 100  & 1/5 & +1.667 & 8.336 \\
PML2 & \(r=.07\) & 1000 & 0/5 &  0.000 & 0.000 \\
V3   & 100 bps   & 20   & 1/8 & -1.099 & 8.788 \\
V3   & 100 bps   & 100  & 0/8 &  0.000 & 0.000 \\
V3   & 100 bps   & 1000 & 0/8 &  0.000 & 0.000 \\
\bottomrule
\end{tabular}
\end{adjustbox}

\end{minipage}
\hfill
% ============================================================
% (b) Size / depth
% ============================================================
\begin{minipage}[t]{0.33\textwidth}
\vspace{0pt}
\centering
\textbf{(b) Order size and depth}
\vspace{3pt}

\begin{adjustbox}{max width=\linewidth}
\begin{tabular}{lrrrrr}
\toprule
\textbf{Market} & \textbf{Size} & \textbf{Shares}
& \textbf{Notional} & \textbf{PnL} & \textbf{Ladder} \\
\midrule
2033169 & .25x & 43.50  & 32.310 & -32.310 & 0.935 \\
2033169 & 1x   & 43.50  & 32.310 & -32.310 & 4.685 \\
2033169 & 4x   & 43.50  & 32.310 & -32.310 & 19.685 \\
\midrule
3082771 & .25x & 150.00 & 42.401 & -42.401 & 0.001 \\
3082771 & 1x   & 396.92 & 112.889 & -112.889 & 0.051 \\
3082771 & 4x   & 781.77 & 224.476 & -224.476 & 0.351 \\
\midrule
3082860 & .25x & 203.11 & 26.185 & -26.185 & 0.006 \\
3082860 & 1x   & 542.98 & 69.563 & -69.563 & 0.031 \\
3082860 & 4x   & 612.98 & 77.863 & -77.863 & 0.000 \\
\bottomrule
\end{tabular}
\end{adjustbox}

\end{minipage}
\hfill
% ============================================================
% (c) Population fee sensitivity
% ============================================================
\begin{minipage}[t]{0.32\textwidth}
\vspace{0pt}
\centering
\textbf{(c) Population fee sensitivity}
\vspace{3pt}

\begin{adjustbox}{max width=\linewidth}
\begin{tabular}{lrrrr}
\toprule
\textbf{Engine} & \textbf{Fee}
& \textbf{Mean ROI} & \textbf{Median ROI}
& \textbf{Flips} \\
\midrule
PML2 & 0   & -6.921 & -1.562 & 0.00\% \\
PML2 & 50  & -6.977 & -1.570 & 2.80\% \\
PML2 & 100 & -7.034 & -1.578 & 2.81\% \\
PML2 & 200 & -7.147 & -1.594 & 15.55\% \\
PML2 & 300 & -7.260 & -1.609 & 39.66\% \\
\midrule
V3 & 0   & +0.626 & +0.114 & 0.00\% \\
V3 & 50  & +0.420 & +0.029 & 5.90\% \\
V3 & 100 & +0.215 & -0.034 & 8.78\% \\
V3 & 200 & -0.196 & -0.209 & 16.39\% \\
V3 & 300 & -0.607 & -0.479 & 21.18\% \\
\bottomrule
\end{tabular}
\end{adjustbox}

\end{minipage}

\vspace{4pt}

\caption{
Execution-cost sensitivity in prediction-market replay.
\textbf{(a)} Fee-induced account feedback under constrained capital; feedback
is native fee-aware PnL minus frozen-path fee accounting.
\textbf{(b)} Order-size and within-execution depth diagnostics for three
preselected PML2 examples; ladder cost is already reflected in execution prices.
\textbf{(c)} Frozen-fill fee sensitivity over 59,356 PML2 and 153,907 V3
episodes; flips denote the fraction of initially positive episodes becoming
non-positive. Fee schedules are experimental assumptions, and the two backend
cohorts should not be interpreted as a paired performance comparison.
}
\label{tab:execution_cost_findings}
\end{table*}

\paragraph{Transaction fees.}
Transaction costs are economically important in prediction markets because
many strategies target small probability mispricings whose gross edge can be
comparable to the trading fee itself. Table~\ref{tab:execution_cost_findings}(a,c)
shows two distinct effects. At ample capital, fees primarily act as a direct
PnL deduction, whereas under tighter cash constraints they can alter subsequent
orders and fills. In the population-scale frozen-fill analysis, increasing
fees progressively erodes returns and converts initially profitable episodes
into non-profitable ones; for V3, the median ROI is already negative at
100 bps despite a positive mean. Thus, post-hoc fee subtraction is adequate
only when account constraints do not change the execution path.

\paragraph{Slippage and market depth.}
Slippage is particularly relevant to prediction markets because displayed
probabilities need not be executable at the desired quantity, especially in
thin or event-driven books. Table~\ref{tab:execution_cost_findings}(b) shows
that larger submitted orders can consume multiple price levels, but the
resulting ladder-cost statistic should not be interpreted mechanically as an
additional strategy loss. For market 2033169, the diagnostic rises from
\(0.935\) to \(19.685\) USDC while total filled quantity, notional, and final
PnL remain unchanged. This occurs because the reference level resets as orders
are split and liquidity is consumed. Execution-cost measures therefore require
an explicit reference price and should be interpreted jointly with realized
fills and account-level PnL.

\section{Discussion}

AlphaOpsBench reframes LLM evaluation in finance by shifting from direct action generation to strategy operationalization. This separation of reasoning from deterministic execution allows us to assess whether models can translate natural-language hypotheses into executable policies while maintaining economic intent. Strategy generation requires interpreting requirements, distinguishing source-specified constraints from model decisions, and mapping abstractions to concrete actions. Failures often arise from silently modifying economic mechanisms, inconsistent parameter completion, or generating unevaluable logic. By explicitly separating source requirements from model choices, our framework identifies specific failure modes rather than aggregating them into opaque financial metrics.

Prediction markets introduce contract-specific challenges beyond price prediction, requiring reasoning about event semantics, information constraints, and settlement mechanics. Our lifecycle-oriented pipeline addresses this through historical execution binding and settlement-aware analysis, using complementary backtesting to capture both aggregate strategy behavior and microstructure effects. However, AlphaOpsBench measures operationalization capability under controlled conditions, not live trading performance. Historical replay cannot fully replicate liquidity dynamics, queue priority, or adaptive counterparty behavior, and the benchmark is constrained by available data and execution interfaces. While future extensions could explore richer multi-contract strategies and adaptive environments, establishing reproducible evaluation baselines is necessary before integrating foundation models into quantitative research workflows.

\section{Conclusion}
\label{sec:conclusion}

We introduced \textsc{AlphaOpsBench}, which treats strategy operationalization---rather than standardized signal generation---as the object of evaluation: source-specified economic requirements are preserved, model-owned decisions are recorded, generated policies are validated at the design, code, and behavioral levels, and research-executable artifacts are linked to historical market, order-book, and settlement evidence. Experiments expose a wide gap between runnable artifacts and faithful operationalization: across five independent full-chain generations, strict canonical acceptance remains sparse on controlled tasks and absent on the 24 preregistered real strategies, and repeated generations vary in model-owned thresholds, sizing, and data choices even when trading direction stays consistent. Historical replay complements this view: large replay coverage implies neither source fidelity nor deployable profitability, financial outcomes depend on execution evidence and modeling assumptions, missing inputs, zero fills, and unresolved settlement must remain distinct from zero return, and fee, capital, and liquidity interventions can alter the subsequent trading path rather than acting as terminal adjustments. These findings support evaluating LLM-based quantitative research by reproducible preservation of economic intent under causal data and market-specific execution and settlement constraints; the present study is limited to one model, a three-month window, and historical execution models that approximate queue state, adaptive counterparties, and live liquidity, and extending models, regimes, and execution evidence is a natural next step.


\newpage
\bibliography{iclr2027_conference}
\bibliographystyle{iclr2027_conference}

\newpage

\appendix
\section{Code and Data Availability}
All code, data, evaluation pipelines, backtesting components, and supplementary scripts necessary to reproduce the reported results will be made publicly available upon acceptance of this paper.

\section{Motivation}
\label{app:motivation}

Recent financial LLM benchmarks have advanced evaluation in QA, quantitative reasoning, alpha discovery, and trading code generation, yet prediction markets remain underexplored. Transferring existing strategy-generation paradigms is insufficient: prediction market strategies rarely reduce to standardized price–volume features or directly compilable numerical rules. Instead, they depend on event and outcome semantics, inter-contract logic, point-in-time information, market lifecycle states, and execution-specific operations. Faithful code generation therefore requires resolving underspecified economic decisions, grounding valid data and entities, preserving temporal causality, and implementing trading semantics. Even executable programs may harbor semantic errors—leaking future information, inverting outcome mappings, or silently altering the intended strategy. These challenges motivate an end-to-end evaluation paradigm that jointly assesses strategy operationalization, code generation, and deterministic behavioral correctness, rather than treating executability as sufficient evidence of successful strategy generation.

\textbf{Traditional financial trading benchmarks.}
Recent financial-AI research has progressed from static question answering to evaluating LLMs as quantitative researchers, strategy designers, and trading agents, spanning formulaic alpha mining~\citep{luo2026alphabench,ding2026alphaeval,chen2026alphasage,wang2026factorminer,wang2026alphamaster}, executable strategy generation and backtesting~\citep{zhang2026alphaforgebench,wang2026backtestbench}, agentic trading systems~\citep{song2026timi,zhang2026finworld,luo2026qfinzero}, and natural-language-to-strategy pipelines~\citep{minara2026}; event-contingent work on merger arbitrage forecasts deal-outcome probabilities but does not address executable strategy construction~\citep{jajal2026globalmerger}. Despite this progress, existing paradigms fall short of prediction-market requirements in three respects: (1)~predefined asset--time abstractions hide the operationalization problem, as most benchmarks assume a fixed universe, standardized features, or a numerical operator vocabulary, whereas prediction-market strategies leave event scope, tradable outcomes, probability sources, logical relations, information timing, and lifecycle policies unresolved; (2)~executability does not imply strategy fidelity, since underspecified decisions resolved implicitly during code generation may silently alter thresholds, outcome directions, sizing rules, or exit conditions, producing syntactically valid code that implements a different economic hypothesis; and (3)~data and execution validity are inseparable from correctness---FINSABER shows that apparent LLM trading advantages deteriorate under broader evaluation~\citep{li2026finsaber}, and prediction markets impose further constraints requiring that heterogeneous external observations be available at decision time and correctly aligned with events and outcomes, while multi-leg, order, lifecycle, and settlement semantics must be preserved. Existing benchmarks evaluate only subsets of these requirements, leaving end-to-end faithfulness of strategy-to-code generation largely unmeasured.

\textbf{AI, blockchain, and prediction markets.}
FinSearchComp evaluates financial search and research, while CryptoAnalystBench examines multi-tool crypto analysis and evidence integration~\citep{hu2026finsearchcomp}. PredictionMarketBench evaluates trading agents via deterministic historical replay~\citep{arora2026predictionmarketbench}; we instead evaluate strategy-program construction from incomplete descriptions. Complementary work has explored event-aware forecasting through on-chain AMM protocols~\citep{jia2026towards}, user-centered transaction understanding with micro-level semantic grounding~\citep{peng2025txsum}, and the economic consequences of cross-chain interoperability~\citep{cao2026price}; however, none addresses the end-to-end operationalization of prediction-market strategies. On Polymarket, outcome tokens represent contingent claims, and strategies may combine trading with collateral splitting, position merging, and redemption of resolved claims under market-specific resolution rules~\citep{{arora2026predictionmarketbench,ng2026price}}. Strategies involving sports, weather, crypto assets, or macroeconomic announcements further require external observations aligned with the correct events and historical datasets. Such evidence constitutes part of the strategy definition rather than interchangeable prompt context, and on-chain observability alone cannot establish its temporal availability or semantic relevance.

\textbf{Limitations of direct strategy generation and our motivation.}
Prediction-market strategies are rarely specified as complete, executable rules, and directly generating code from coarse descriptions forces an LLM to simultaneously resolve several interrelated challenges: underspecified descriptions that omit target outcomes, probability models, thresholds, sizing, order types, or exit rules, where silent completion may alter the original economic hypothesis; event and temporal alignment, requiring external signals to correspond to the correct events and outcomes and be available at decision time, otherwise seemingly plausible code may harbor semantic or point-in-time leakage; prediction-market trading semantics, where generated programs may invert outcomes, omit necessary steps, or mis-handle lifecycle, settlement, and multi-leg constraints; and the fundamental gap between executability and fidelity, as code may compile and backtest successfully yet implement behavior divergent from the intended strategy. Consequently, direct generation may produce strategies that are runnable but semantically invalid. Rather than treating strategy generation as a one-shot natural-language-to-code task, we benchmark the complete translation from incomplete economic ideas to auditable, executable trading strategies, requiring that a valid implementation preserve the intended strategy, respect event and outcome semantics, use only point-in-time information, and satisfy deterministic behavioral constraints. This motivates end-to-end benchmarking of prediction-market strategy operationalization and code generation, where success demands not merely executable code but faithful and reproducible strategy behavior.

\section{detailed construction of our data}
\label{data_construction}
\paragraph{Polymarket Infrastructure and Data.} Polymarket operates as a blockchain-based prediction market where users trade ERC-1155 conditional tokens (built on the Gnosis Conditional Token Framework) on the Polygon mainnet. It employs a hybrid decentralized exchange architecture: order matching occurs via an off-chain Central Limit Order Book (CLOB), while settlement, token splitting/merging, and redemptions are executed via on-chain smart contracts. This setup enables highly reproducible quantitative research. Market metadata, resolution sources, and historical prices are accessible via public APIs, while on-chain Polygon records provide an immutable, verifiable ledger of trades and settlements, enabling full lifecycle reconstruction. Since complete limit order book (LOB) snapshots are not natively stored on-chain, we continuously ingest and archive real-time market depth directly from the CLOB data stream. 

\paragraph{Institutional Evolution and Study Window.} Polymarket's technical and institutional infrastructure has undergone significant evolution. Key transitions include shifts in oracle resolution mechanisms (e.g., UMA Optimistic Oracle, NegRisk adapter), fee structures (from independent modules to protocol-level fees), and core smart contract generations (upgrading from the legacy CTF Exchange to CTF Exchange V2 with new signature and collateral architectures). Merging data across these disparate eras conflates strategy performance with shifting contract semantics, oracle behaviors, and fee regimes, introducing severe non-stationarity. To mitigate this, we restrict our primary study window to June 1 through September 1, 2026. This three-month interval falls strictly after the V2 migration, ensuring a consistent market, trading, and settlement environment, thereby balancing sample size with institutional stationarity.

\paragraph{Heterogeneity of Prediction Market Contracts.} Conventional trading universes such as equities or cryptocurrencies share standardized price, volume, and return representations; prediction markets do not. Polymarket spans politics, macroeconomic releases, sports, crypto assets, weather, and entertainment, with each market defining its own outcome space, resolution source, trading horizon, and settlement rules. Contracts therefore differ not only in subject matter but in what counts as a valid signal, which external observations they require, how outcomes map to tradable tokens, and when information arrives. Strategy generation must recover this event structure and its temporal alignment before emitting trading logic compatible with the market's lifecycle and execution constraints---strictly harder than reusing one formula across assets.

\subsection{Polymarket Dataset}
\label{app:polymarket_dataset}

All statistics refer to the half-open UTC window [2026-06-01 00:00, 2026-09-01 00:00), spanning 92 days and 2,208 hourly partitions. We distinguish the cleaned execution--settlement base, the markets in which trading and settlement evidence intersect, and the strategy-specific subsets consumed by experiments; settlement figures use a census saved at 2026-09-12 00:00 UTC, and statements about the base are not claims that every strategy was backtested over it.

\paragraph{Execution stream.} Trades are reconstructed from \texttt{OrderFilled} events on Polygon, read jointly from PostgreSQL and ClickHouse, deduplicated, time-aligned on block time, linked to market and token identities, and reduced to an aggressor-side tape (Table~\ref{tab:ds_exec}); passive legs are excluded so that the passive side of a match is not counted twice. Cleaning normalizes on-chain identifiers losslessly, deduplicates on (chain, contract, transaction hash, log index) with cross-source conflicts logged, links identities via market--token maps with protocol position relations as fallback, and recovers trade direction from contract-version-specific fields rather than price movement. Market and token totals (1,348,555 and 1,978,115) are deduplicated across months and are not column sums. Strategy inputs such as volume and rolling prices are computed from historical prefixes of this tape; trades are never compressed into OHLCV bars before execution modeling. The catalog spans all 2,208 hourly partitions and is stored as market-clustered Parquet.

\paragraph{Market composition.} All traded markets join to metadata, with identities from canonical Gamma (997,397), legacy combo (179,090), placeholder (99,836), and protocol-structure (71,874) sources; we therefore describe the universe as market identities with linkable conditions. Sports (47.07\%) and crypto (25.62\%) dominate, so per-category sample sizes differ by up to two orders of magnitude. 

\paragraph{Resolution and settlement.} Settlement is reconstructed from lifecycle events---requests, proposals, disputes, rulings, adapter and module operations---rather than final labels; June--August contain 13,531,251 such records, and a single market typically generates several, so these are not settled-market counts. Closure flags or a ruling at one oracle layer alone do not establish a redeemable payout. At the census cutoff, 1,323,315 of 1,348,555 traded markets (98.13\%) carry consumable payout evidence, of which 1,310,720 settled before September 1; most of the 25,240 remainder are unresolved at the cutoff rather than collection failures (Table~\ref{tab:ds_settle}). A stricter population of 1,322,654 markets additionally requires one in-window fill strictly before settlement. Payouts are not uniformly binary: 972,377 markets settle via CTF evidence and 350,938 via combo or module evidence, including 7,342 equal-refund and 184 other fractional vectors, retained as numerator--denominator pairs rather than coerced to winner-1/loser-0. Consumable denotes payout evidence plus simulated-ledger checks, not on-chain redemption or profit.

\paragraph{Limit order book.} Depth is archived separately from fills as full-depth L2 state covering all quoted price levels: for each replay window, per-level quantities for both outcome tokens are reconstructed from published baseline snapshots and ordered increments, with YES and NO books kept separate and exchange event time separated from collector reception time. The archive covers 2,093 of 2,208 hours (94.79\%) and, cumulatively, a majority of traded conditions (Table~\ref{tab:ds_lob}); the 115 unobserved hours concentrate in August. Hourly presence implies neither intra-hour continuity nor two-sided depth for both tokens, so per-token availability is checked before replay and gaps are retained as explicit missingness.

\paragraph{Data quality.} Activity is skewed---35.47\% of markets have a single aggressor fill and the per-market median is four---so panel selection, not raw market counts, determines which markets support continuous decision analysis. Two properties are disclosed: 175,732,742 fills (95.70\%) lack a usable intra-block index and are replayed under deterministic degraded ordering without queue-priority claims, and 60,365,248 records (32.87\%) show source price inconsistent with the amount-to-size ratio, with both fields retained. Post-cleaning checks find no non-positive sizes, out-of-range prices, or unresolved field conflicts, and the 148,903 fills not preceding their market's settlement are handled by lifecycle tradable-window logic. Within these disclosed limits and the explicit liquidity and execution models of the backtesters, the base supports reproducible backtesting.

\begin{table*}[t]
\centering
\scriptsize
\setlength{\tabcolsep}{3pt}
\begin{minipage}[t]{0.32\textwidth}
\centering
\caption{Execution cleaning funnel. The first row counts redundant reads across stores, not on-chain trades.}
\label{tab:ds_exec}
\begin{tabular}{l r}
\toprule
Stage & Records \\
\midrule
Physical reads & 681,516,938 \\
Dup.\ reads removed & -187,063,483 \\
Out-of-window & -438,963 \\
Deduped events & 494,014,492 \\
Passive legs & -310,388,628 \\
Unlinkable & -71 \\
\midrule
Aggressor fills & 183,625,793 \\
\bottomrule
\end{tabular}
\end{minipage}\hfill
\begin{minipage}[t]{0.30\textwidth}
\centering
\caption{Settlement census at 2026-09-12 00:00 UTC.}
\label{tab:ds_settle}
\begin{tabular}{l r}
\toprule
Condition & Markets \\
\midrule
Traded & 1,348,555 \\
Payout evidence & 1,323,315 \\
\quad settled $<$ Sep 1 & 1,310,720 \\
No evidence & 25,240 \\
\bottomrule
\end{tabular}
\end{minipage}\hfill
\begin{minipage}[t]{0.32\textwidth}
\centering
\caption{Hourly LOB coverage: hours with depth over calendar hours, and within-month distinct conditions.}
\label{tab:ds_lob}
\begin{tabular}{l r r}
\toprule
Month & Hours & Conditions \\
\midrule
June & 717/720 & 141,506 \\
July & 744/744 & 520,421 \\
Aug & 632/744 & 841,468 \\
\midrule
Total & 2,093/2,208 & -- \\
\bottomrule
\end{tabular}
\end{minipage}
\end{table*}

\section{Construction of \textsc{AlphaOpsBench}}
\label{app:benchmark_construction}

\textsc{AlphaOpsBench} tests whether an LLM can convert an incompletely specified prediction-market trading idea into an executable strategy without silently altering the economic hypothesis under test. This appendix documents its construction and validation pipeline, which comprises four stages: (i) collecting real-world strategy sources with full provenance; (ii) canonicalizing semantically equivalent strategies and compiling them into source-preserving tasks; (iii) constructing controlled diagnostic variants that span specification-completeness levels; and (iv) generating strategies under Direct and Staged protocols, followed by deterministic compilation and validation. Failed, unsupported, ambiguous, and unverified cases are retained rather than filtered, so that design failures remain separable from those caused by missing data, unsupported execution semantics, ambiguous sources, or limited validation coverage.

\subsection{Real-World Strategy Sources}
\label{app:real_world_strategy_sources}

The real-world track comprises 581 \texttt{SourceRecord}s that preserve coarse public trading hypotheses together with their provenance: original source text, document- and text-level hashes, source location, source family, and downstream task lineage. Records enter the corpus without any assumption of benchmark readiness; reusable execution components and problematic source material are retained here and resolved in later stages. The collected corpus exceeds the final task population by design, since semantic deduplication, capability checks, and task materialization are deferred to subsequent stages. The corpus, acquisition prompts, and lineage maps will be released upon acceptance.

\paragraph{Hand-curated guide (128 records).} A manually curated set of coarse Polymarket strategies covering structural arbitrage, logical consistency, market making, order-book microstructure, external probability anchors, domain forecasting, wallet signals, and portfolio risk. Records are kept at the level practitioners describe them: the economic mechanism is stated while lookback windows, thresholds, sizing, and exits often remain open.

\paragraph{Web-research catalogs (119/120/112 records).} ChatGPT, Qwen, and Grok each produced an independent catalog under the common acquisition specification described below, retaining 119 of 284 and 120 of 265 examined candidates respectively, with Grok contributing 112 retained records. Independent generation supplies source diversity; mechanism-level canonicalization later decides whether a record contributes a distinct strategy or only a variant of an existing one.

\paragraph{Academic literature (102 records).} Papers and working papers on prediction markets, forecasting, microstructure, sports, macroeconomics, and weather were reviewed, and 102 records whose excerpts, data requirements, decision logic, trading actions, and exit/settlement semantics fit the benchmark schema were retained.

\paragraph{Source-grounded acquisition protocol.} The web-research runs treat the LLM as a source-discovery and evidence-synthesis agent rather than a strategy inventor. The agent first verifies the current Polymarket environment---order types, fees, incentives, position operations, negative-risk mechanics, token structure, resolution, and public interfaces---from official documentation and repositories, then builds a candidate pool from documentation, code repositories, academic papers, on-chain analyses, wallet studies, and technical discussions, reading each source rather than relying on search snippets. A candidate is eligible only if it identifies an edge source, an observable signal or data requirement, the traded market or outcome, a position or order construction, an entry condition, an exit or settlement rule, and a principal risk; external venues may inform the signal, but the profit-generating trade must occur on Polymarket. Deduplication operates on economic mechanism rather than wording: substitutions of asset, league, horizon, threshold, or YES/NO orientation do not define new strategies. If verified candidates fall short of the target catalog size, the agent reports the shortfall instead of fabricating entries.

\begin{sourceexample}{Representative Rough Strategy Record}
\textbf{Binary complete-set buy arbitrage.} ``If \(ask_{\mathrm{YES}} + ask_{\mathrm{NO}} + \mathrm{fees} + \mathrm{slippage} < 1\), buy equal quantities of YES and NO using FOK orders, then merge the complete set or hold to settlement.'' The record fixes the economic mechanism---complete-set payoff parity---while leaving implementation choices open. The gap between such a rough hypothesis and a fully specified trading program is the object of evaluation.
\end{sourceexample}

\paragraph{Acquisition prompt.} The prompt used in the collection runs additionally fixes source priorities, evidence grades, current-validity checks, and audit requirements, and is reproduced in full in the supplementary material; its eligibility core is summarized below.

\begin{promptbox}{Source-Grounded Polymarket Strategy Acquisition Prompt}
\textbf{Role and objective.} Act as a research agent in prediction markets, market microstructure, on-chain analysis, and Polymarket; compile a catalog of genuinely distinct strategies supported by direct, verifiable public evidence, building a broad candidate pool before selection.

\textbf{Polymarket-only execution.} Retained strategies must place orders, hold or manage positions, or perform split, merge, conversion, or redemption on Polymarket; external platforms and datasets may serve only as signal sources. Renaming another venue's strategy does not qualify it: adaptations must address binary or categorical payoff, execution, liquidity, fees, settlement, and resolution semantics.

\textbf{Eligibility.} Each strategy states (1) its economic or informational edge, (2) the observable signal or required data, (3) the market or outcome traded, (4) the position or order construction, (5) the entry condition, (6) the exit, unwind, merge, conversion, redemption, or settlement rule, and (7) the principal execution or model risk. Directives such as ``use news,'' ``follow smart money,'' or ``apply machine learning'' are ineligible until converted into a concrete, sourced mechanism.

\textbf{Procedure and adaptation.} Verify the live Polymarket environment from official documentation before collection, and open and read each candidate's source. A cross-domain adaptation is admissible only with an accessible original source, an explicit mapping to a Polymarket outcome, and a concrete signal and trading rule.

\textbf{Deduplication, evidence, anti-fabrication.} Candidates sharing edge source, primary signal, payoff construction, entry and exit logic, and profit mechanism are duplicates; asset, league, horizon, threshold, orientation, and naming changes are not new strategies. Every retained strategy carries an origin label, an evidence grade measuring source support rather than expected profit, and a current-validity status. Do not invent statistics or strategies: report a shortfall rather than pad the catalog.
\end{promptbox}

\subsection{Rules in Polymarket}
\label{sec:polymarket-rules}

Traditional quantitative trading typically relies on a homogeneous buy/sell/hold action space over standardized assets. In contrast, Polymarket requires strategies to operate on heterogeneous outcome tokens (e.g., YES or NO) rather than undifferentiated asset shares. A position represents a conditional claim on a specific event outcome, making resolution rules and oracle sources intrinsic to the instrument definition. Consequently, selling a YES token is not a generic short position but the disposal of a specific contingent claim.

Beyond directional orders, the venue imposes complex execution attributes and protocol-level operations. Orders are bound to specific outcome tokens and subject to time-in-force constraints such as Fill-and-Kill (FAK) and Fill-or-Kill (FOK), which dictate immediate execution conditions without guaranteeing atomic multi-leg fills. Polymarket also exposes collateral and token operations outside the standard central limit order book (CLOB). Strategies can split collateral into matching YES and NO positions, merge them to recover collateral, or redeem winning tokens post-resolution. For mutually exclusive multi-outcome events, the negative-risk adapter allows converting a NO token on one outcome into YES tokens on others, provided the token universe and event mapping are strictly verified. These mechanics expand the strategy space beyond single-market directional trades to include complete-set arbitrage, multi-outcome baskets, negative-risk parity, and cross-market payoff-consistency constructions. We construct our action vocabulary from 581 source-preserving strategy records, encompassing hand-curated guides, LLM-generated catalogs, and academic literature. After mechanism-level deduplication based on payoff source, primary signal, and state transitions, we distill 106 canonical strategy families and 27 shared execution components. This yields a comprehensive action schema covering directional trades, quote management, grouped order legs, and protocol operations.

\subsection{Detailed Strategy Generation}
\label{app:strategy_generation}

The generation stage translates each source-preserving task into an explicit trading policy before historical execution. Unlike factor benchmarks that map natural-language queries to signals over OHLCV bars, prediction-market policies depend on outcome tokens, event semantics, execution types, position state, and protocol actions.

\paragraph{Common envelope and policy form.} Both arms receive an identical task envelope---strategy description, source-derived requirements, factor catalog, execution-data policy, action capabilities, and research environment---and differ only in how the task is operationalized. A design names a logical target (market role and outcome, e.g.\ primary/YES, with physical tokens bound at runtime), inputs drawn from registered factors, declared historical transformations, or supported formulas (account variables are runtime-bound, not market features), and an ordered rule list $r_i=(c_i,a_i)$; evaluation executes only the first rule whose condition holds, $i^\star=\min\{i:c_i(x_t)=1\}$. An order action specifies at least $(\mathrm{side}, \mathrm{price}, \mathrm{quantity}, \mathrm{outcome}, \mathrm{TIF})$; multi-leg, cancel/replace, and split/merge requirements keep their native action types and are never collapsed into a single BUY or SELL.

\paragraph{Direct and Staged protocols.} Direct asks one invocation to resolve the open economic choices and return design and code jointly, $T \xrightarrow{\mathrm{LLM}} (D,C)$. Staged first elicits the design alone, $T \xrightarrow{\mathrm{LLM}} D$, applies deterministic host-side schema, source-requirement, capability, and market-domain checks plus lowering to the shared policy form without further LLM calls, and then sends only the validated design $\widetilde D$ to a second invocation for implementation, $\widetilde D \xrightarrow{\mathrm{LLM}} C$. Staged thus tests whether separating what to implement from how to encode it improves faithfulness.

\paragraph{Source preservation and completeness levels.} Requirements stated by the source are binding; parameters the source leaves open (thresholds, windows, sizing, exits) may be chosen by the model, while missing historical coverage licenses neither invented inputs nor alternative APIs---the design records the extra requirement or abstains, and declared-but-unused inputs do not satisfy a requirement. Diagnostic tasks vary decision ownership at fixed mechanism identity: L1 supplies mechanism, rules, and parameters; L2 opens selected choices; L3 supplies only the objective and interface.

\paragraph{Prompts and illustration.} Both arms share a prompt assembled from the task envelope, the design schema, and an interface example that fixes output syntax without supplying the economic answer; the system instruction requires preserving the source strategy, filling only open choices, and keeping first-match rule semantics. Direct must return design and code in one JSON object; the Staged designer returns the design only. A representative design maps registered inputs (acceleration, last price, cash, position) to ordered rules, e.g.\ buy YES as taker when acceleration $>0$ and price $<0.8$ with size $\min(0.01\,\mathrm{cash}/p,\,10)$, sell the position when acceleration $<0$, and hold otherwise.

\subsection{Code Generation}
\label{app:code_generation}

The validated policy is translated into an executable candidate under a restricted Python interface: the model implements the comparisons, arithmetic, rule order, and requested actions, while market binding, account state, execution, settlement, and data access are supplied by the common runtime. Every candidate contains exactly one class, \texttt{GeneratedActionStrategy(ActionStrategyV1)}, whose economic logic lives solely in \texttt{decide(self, snapshot)}; the language excludes imports, network, file-system, and wall-clock access, helper methods, persistent model state, and strategy interpreters, and injects \texttt{Decimal}, action enums, and the base class from the host. Factor and asset accessors must quote literal registered identifiers, making each candidate's data dependencies statically auditable.

At each decision time the candidate observes $S_t=(X_t,A_t,M_t)$---registered market and derived factors, causal account state, and runtime-bound metadata---and implements $\pi_C: S_t \rightarrow \mathcal{A}$ over supported action plans; visibility of historical records, order arrival, fills, and cash updates are decided by the environment, not the model. Ordinary orders are submitted through a single helper, \texttt{submit\_order(side, price, quantity, outcome, market\_ref, rule\_index, execution)}, which binds task, market, token, and order identities mechanically while leaving signal conditions and sizing in the generated code; absent rules return a hold, and grouped orders, cancel/replace, quote pairs, position and protocol operations use typed action constructors rather than rewritten ordinary orders.

In the Direct arm the code shares the design's response and is never repaired by a second model, so a malformed candidate remains a failed sample. In the Staged arm the coder receives only the validated design and the registered factor identifiers, is prohibited from redesigning the strategy or revisiting the source, and must reproduce the design's arithmetic, comparisons, action parameters, and first-match order exactly.

All post-generation processing is deterministic: parsing, mechanical schema alignment, materialization into the shared action-policy form, and loading into the restricted runtime, followed by static checks of the candidate language and literal dependencies, source-requirement checks where machine-checkable evidence exists, and behavioral tests on controlled snapshots covering action direction, threshold boundaries, outcome binding, missing-input handling, and position-aware follow-ups. A parsed or loaded candidate is therefore not presumed semantically correct, and failures, unverified requirements, and unsupported capabilities are retained as distinct outcomes. The candidate only requests actions: historical time, latency, arrival, liquidity, capital, fees, fills, account evolution, and settlement belong to the shared backtester, so identical economic logic may yield different outcomes under trade-tape and order-book execution models, and execution success is never conflated with semantic correctness.

\subsection{Backtesting framework}
\label{backtest}
\paragraph{Market eligibility and point-in-time data construction.}
Before running either backtesting engine, we construct a market eligibility set rather than applying every strategy to every Polymarket market. Each candidate market must have a stable mapping among the market identifier, condition identifier, outcome token identifiers, and outcome labels. We then screen markets using historical activity and data quality, including observed trade count, traded notional, active trading intervals, lifecycle status, and the availability of causally ordered execution records in the target window. Markets with no meaningful trading activity, missing token mappings, unresolved lifecycle identity, or insufficient historical execution coverage are excluded from the strategy universe. This screening criterion is a sample-construction rule rather than a guarantee of execution capacity: a market can be active enough to enter the research universe while still producing a partial fill or no fill for a particular order. We therefore keep market eligibility, execution eligibility, and realized fill status as separate fields in the replay output.

Both engines use point-in-time inputs. A strategy decision at time $t$ can only consume records whose availability time is no later than $t$. Future trades, future book states, market-resolution labels, and post-decision settlement information are excluded from the signal input. The market and token identities are bound to the execution request, so a trade or order from another condition, outcome token, or market cannot be reused accidentally. The Alpha layer generates typed actions and records the strategy decision, while the shared backtesting backend owns historical time advancement, order admission, matching, capital reservation, account state, settlement, and financial reporting.

\paragraph{Fill-only V3.} V3 is a trade-tape execution model for strategies whose logic requires neither order-book depth nor queue position. It replays historical trade prints---market and token identity, outcome, block time, price, quantity, aggressor side, and lineage---indexed by market and token so that each order scans only its causal interval. An order issued at signal time \(t_s\) with modeled latency \(\ell\) arrives at \(t_a = t_s + \ell\) and may execute only within \(\mathcal{W}(o) = [t_a,\, \min(t_a + H, t_d)]\), against same-market, same-token prints satisfying its side and limit-price constraints; if the signal was itself triggered by a historical print, that print is excluded from executable capacity, so one event cannot both generate and fill the trade.

\medskip\noindent\textbf{Shared and synthetic capacity.} In source-confirmed replay, an eligible print \(j\) with quantity \(q_j\) contributes \(c_j = \rho q_j\) at declared participation rate \(\rho\), and a shared liquidity ledger tracks residual capacity \(c_j^{\mathrm{rem}} = \max\bigl(0,\, c_j - \sum_{o' \prec o} q^{\mathrm{fill}}_{o',j}\bigr)\), so overlapping orders compete and no observed volume is reused. The constrained experiments instead freeze the pre-arrival tape to estimate the same-side flow rate \(r_{\mathrm{side}} = V_{\mathrm{side}}^{\mathrm{pre}} / \max(\Delta_{\mathrm{pre}}, 1\,\mathrm{s})\); each one-second cell \(k\) in the horizon then contributes \(C_k = r_{\mathrm{side}} \Delta t_k\, \rho\, m\), giving \(C(o) = \sum_{k \in \mathcal{W}(o)} C_k\) under stress multiplier \(m\). The reported panel uses \(H = 60\) s, \(\rho = 1\), \(m = 1\); \(m = 0.5\) and \(m = 2\) define stress variants outside it and are declared settings, not calibrated quantiles. Synthetic capacity is labeled modeled liquidity and never reported as an observed fill.

\medskip\noindent\textbf{Execution price and fills.} Prices derive from a latent midpoint and spread inferred from the pre-arrival tape: the midpoint is the mean of the most recent buy--sell pair, or an exponentially smoothed trade price (weight 0.35 on the newest print) when no pair exists; the spread is the observed buy--sell gap or a Roll-type estimate, floored at one tick. A taker buy with limit \(L\) executes at \(\min(L,\, m_t + s_t/2)\) and a taker sell symmetrically at \(\max(L,\, m_t - s_t/2)\), separating execution price from signal price. Filled quantity is \(Q^{\mathrm{fill}} = \min\{Q,\, C^{\mathrm{rem}}(o)\}\); native FAK leaves partial fills and native FOK is all-or-none, while the constrained panel remaps IOC and FOK to FAK, a simplification reported separately from native behavior.

\medskip\noindent\textbf{Scope.} V3 targets broad, low-cost evaluation of taker-style strategies: it models no queue position, post-only maker semantics, persistent resting orders, or cancel/replace, which are routed to PML2's L2 replay. Execution uncertainty enters only through causal flow, participation, arrival, price, and capacity assumptions. Constrained runs fix fees at zero, mark fills as synthetic-arrival liquidity under an uncalibrated tape proxy, and value terminal positions at the last observed midpoint; returns are therefore controlled low-fidelity execution outcomes, not estimates of on-chain fills or deployable performance.

\paragraph{PML2 trade-tick and LOB replay.}
PML2 is the execution framework for strategies that depend on visible depth, spread, executable size, maker liquidity, queue position, or stateful order lifecycle. Its input is an archive of historically collected LOB snapshots, book-frame updates, price-level changes, and trade ticks. The archive provider restores only the market intervals reached by the strategy's actual order arrivals. Events are replayed in exchange-event order and bound to source manifests, coverage receipts, and file hashes. A restored book is accepted only when the market and token identities match, the book is two-sided, the snapshot is non-crossed, the data is within the configured freshness bound, and no unresolved gap is present.

The PML2 point-in-time surface derives execution factors from a single accepted book frame for each outcome token. These factors include best bid and ask, depth-weighted prices, visible executable capacity, estimated slippage, fee information, book freshness, gap status, and market status. YES and NO books are maintained as separate token frames. Even when their events share an exchange timestamp, the implementation does not treat them as an atomic cross-outcome snapshot. Consequently, a multi-leg or complete-set strategy cannot obtain atomicity merely from simultaneous timestamps.

PML2 uses an event-driven replay session. A strategy action is submitted at its decision time, subjected to the configured entry latency, and admitted at a modeled venue-arrival time. For an aggressive BUY, the matcher walks eligible ask levels in price order; for an aggressive SELL, it walks eligible bid levels. At each level, only the declared fraction of visible size is executable, and the engine records the level, consumed size, execution price, source event identifiers, book timestamp, liquidity role, and any modeled fee. If the visible depth is insufficient, the order is treated according to its time-in-force. FOK rejects the order when the full quantity cannot be committed. FAK and IOC retain any executed quantity and cancel the remainder. GTC and GTD orders can leave a residual resting order after the immediately executable portion has been processed.

Resting orders are represented by an explicit queue ledger at each asset, side, and price level. The initial queue ahead is estimated from visible external size and the selected PML2 profile. Subsequent OrderFilled trade ticks consume queue capacity in price-time order. In profiles that enable it, unexplained LOB decreases can also reduce modeled queue ahead using a declared cancellation policy. These queue positions are therefore model-based reconstructions from public LOB and trade evidence; they are not observed private queue identities. The engine also tracks cancel and replace latency, order expiration, partial states, reservations, and order-group outcomes. Cross-decision cancellation or replacement is accepted only when the target order was introduced by an earlier decision in the same hash-bound strategy trace. Unsupported position, protocol, or kill-switch actions fail closed at the trace adapter rather than being silently converted into ordinary orders.

PML2 maintains an independent account for each strategy candidate. Before execution, the portfolio reserves the required cash or token inventory. Each match updates the reservation and account journal, while explicit settlement receipts, portfolio marks, or market-redemption events are processed chronologically after the corresponding fill. An order with a stale, missing, crossed, or gapped book is not converted into an ordinary no-fill observation: it receives a data-readiness or rejection state, preserving the distinction between unavailable execution evidence and genuine lack of liquidity. 

\subsection{Controlled Diagnostic Tasks}
\label{app:controlled_diagnostics}

We construct 36 controlled diagnostic tasks from twelve base tasks spanning
six prediction-market mechanisms: complete-set trading, directional
probability changes, inventory-constrained quoting, trade-flow following,
external-signal residuals, and settlement or lifecycle actions.
Each base task is instantiated at three specification levels, L1--L3,
which differ in the ownership of economic decisions rather than in an
assumed empirical ordering of difficulty.

At L1, the task fixes the data contract, factor definition, lookback
window, threshold, direction, order policy, sizing, exit condition, and
no-trade rule, so the primary challenge is faithful implementation.
L2 retains the economic mechanism and binding source requirements while
leaving selected parameters or implementation choices open. L3 specifies
the objective and admissible interfaces but delegates most of the
economic design to the generator. The resulting levels therefore measure
\emph{specification completeness}; we do not assume that L3 must be
empirically harder than L1.

External-signal diagnostics use explicitly defined test mappings rather
than calibrated forecasts of fair probabilities. Lifecycle tasks may
legally remain inactive when initialized without the inventory required
by the source mechanism. Constructed common states provide inventory only
where the diagnostic scenario itself requires it; this does not authorize
inventory injection into historical replay accounts. L1 tasks can be
checked against explicit reference actions. For L2 and L3, a reference
implementation represents one admissible completion of the task rather
than a unique economic solution or a profitability upper bound.

\begin{table*}[t]
    \centering
    \small
    \setlength{\tabcolsep}{5pt}
    \caption{
    Twelve controlled base tasks, each instantiated at L1--L3.
    The listed checks characterize the intended diagnostic mechanism and
    do not define a unique solution for underspecified L2 or L3 tasks.
    }
    \label{tab:controlled_diagnostic_tasks}
    \begin{tabular}{lll}
        \toprule
        Base task & Description & Principal checks \\
        \midrule
        \texttt{complete\_set}
        & Complete-set discount
        & Two BUY legs; same condition; cash/capacity bounds \\

        \texttt{directional\_move}
        & Directional momentum
        & Time-based probability return; entry and inventory exit \\

        \texttt{inventory\_quotes}
        & Inventory-constrained quoting
        & Two-sided quotes; cash and inventory constraints \\

        \texttt{trade\_flow}
        & Share-flow following
        & Registered signed flow; time window and threshold \\

        \texttt{complete\_set\_premium}
        & Complete-set premium
        & Two SELL legs from inventory; no naked short \\

        \texttt{directional\_reversion}
        & Directional reversal
        & Reverse direction; boundary and inventory exit \\

        \texttt{inventory\_microprice}
        & Microprice inventory quoting
        & Depth-weighted center and inventory skew \\

        \texttt{trade\_flow\_normalized}
        & Normalized flow following
        & Signed shares / total shares; zero denominator \\

        \texttt{external\_crypto}
        & External spot residual
        & Causal spot return and specified diagnostic mapping \\

        \texttt{external\_cpi}
        & External CPI residual
        & First-release value; causal availability and mapping \\

        \texttt{lifecycle\_redeem}
        & Resolution redemption
        & Verified resolution and eligible existing inventory \\

        \texttt{lifecycle\_cancel}
        & Lifecycle cancellation
        & Closed market; existing open orders; typed cancellation \\
        \bottomrule
    \end{tabular}
\end{table*}

\paragraph{Evaluator audit.}
Source-fidelity checks, preservation of source-given economic slots,
completion of model-owned slots, and observed behavioral agreement are
reported separately. Reference and mutation probes are additionally used
to inspect whether the diagnostic checker recognizes known-valid and
known-invalid behaviors. These probes reveal blind spots involving target
outcomes and multi-step actions, together with overly restrictive behavior
at two inventory-boundary cases. We therefore treat the local behavioral
checker as one source of evidence rather than as an independently validated
complete semantic oracle; canonical source-aware evaluation remains a
separate layer.

% \tcbinputlisting{
%     promptbox,
%     title={Source-Grounded Polymarket Strategy Acquisition Prompt},
%     listing file={appendix/prompts/polymarket_strategy_acquisition_prompt.txt}
% }

\section{Detail Analysis}

\subsection{Worked Example: Operationalizing a Volatility-Adaptive Market-Making Strategy}
\label{app:worked_example_c029}

This section illustrates the complete operationalization process using a
representative real-world strategy from the benchmark.
The example is intentionally chosen to expose the distinction between a coarse
economic hypothesis, the choices made by the language model, and the behavior
that is ultimately evaluated by the historical execution environment.
It therefore serves as a diagnostic example rather than as a demonstration of
profitable market making.

The source strategy describes a volatility-adaptive market-making mechanism.
Its central hypothesis is that quote aggressiveness should vary with short-term
uncertainty in the market-implied probability: when probability volatility is
high, a liquidity provider should quote more conservatively; when volatility is
low, quotes may be tightened subject to a minimum edge and an inventory
constraint.
The source further specifies maker-oriented execution through post-only orders,
but leaves the numerical volatility definition, thresholds, quote offsets,
position size, and detailed exit policy unspecified.
These quantities remain model-owned decisions under our task
ownership protocol.

\begin{strategybox}{Source Strategy: Belief-Volatility Adaptive Market Making}

Estimate short-horizon volatility of the market's implied probability and
widen spreads, reduce order sizes, or shorten quote lifetime when volatility
rises.
When the market is stable, narrow quotes while maintaining a minimum expected
edge and respecting inventory constraints.
The intended source of return is selective spread capture rather than
directional exposure.
The strategy is intended to use resting, post-only orders, with time-bounded
quotes when appropriate.

\end{strategybox}

\paragraph{From economic intent to an executable design.}
The source specifies the qualitative response to volatility but does not define
a unique quantitative implementation.
In the Direct condition, the model resolves these open choices by using the
absolute one-step change in market-implied probability as its volatility proxy,
denoted by
\[
v_t = |\Delta p_t|.
\]
The generated design trades the YES outcome only and introduces two regimes.
A high-volatility regime is activated when
\[
v_t > 0.02
\quad\text{and}\quad
s_t < 0.05,
\]
where \(s_t\) is the contemporaneous bid--ask spread.
A low-volatility regime is activated when
\[
v_t \leq 0.02
\quad\text{and}\quad
s_t < 0.02.
\]

The model further chooses regime-dependent quote offsets and order-size caps.
Table~\ref{app:worked_example_c029} summarizes the resulting policy.

Here \(C\) denotes the recent executable-capacity proxy available to the
strategy.
The generated policy therefore instantiates the source idea with explicit
thresholds, order prices, and sizing rules that were absent from the original
description.

\paragraph{Generated decision logic.}
The corresponding executable policy can be summarized by the following
decision rule.
We omit framework-specific boilerplate and retain only the economic logic used
at each decision point.

\begin{policybox}{Generated Direct Policy (Simplified)}

\textbf{Inputs.}
Let \(v_t\) denote the short-horizon probability-change proxy,
\(s_t\) the bid--ask spread,
\(C_t\) the recent executable-capacity estimate,
and \(I_t\) the current YES inventory.
If any required input is unavailable, the strategy returns \textsc{HOLD}.

\medskip
\textbf{High-volatility regime.}
If
\[
v_t > 0.02
\qquad\text{and}\qquad
s_t < 0.05,
\]
the strategy sets
\[
P_t^{buy}=\max(0.01,bid_t+0.05),
\qquad
P_t^{sell}=\min(0.99,ask_t-0.05),
\]
with
\[
Q_t^{buy}
=
\min\left(
\frac{0.005\,Cash_t}{P_t^{buy}},
0.5C_t,
5
\right),
\]
and
\[
Q_t^{sell}
=
\min(I_t,0.5C_t,5).
\]

\medskip
\textbf{Low-volatility regime.}
If
\[
v_t \leq 0.02
\qquad\text{and}\qquad
s_t < 0.02,
\]
the corresponding rules are
\[
P_t^{buy}=\max(0.01,bid_t+0.01),
\qquad
P_t^{sell}=\min(0.99,ask_t-0.01),
\]
\[
Q_t^{buy}
=
\min\left(
\frac{0.01\,Cash_t}{P_t^{buy}},
C_t,
10
\right),
\qquad
Q_t^{sell}
=
\min(I_t,C_t,10).
\]

\medskip
In both regimes, the generated policy requests two-sided
\textbf{maker, post-only, GTC} orders.
If neither regime is active, the strategy returns \textsc{HOLD}.

\end{policybox}

\paragraph{Source-fidelity analysis.}
Although the generated program is executable, its operationalization is not
fully equivalent to the source strategy.
This example exposes three forms of semantic drift that motivate our
source-aware evaluation.

First, the source refers to short-horizon probability volatility, whereas the
generated design uses the absolute one-step probability change,
\[
|\Delta p_t|,
\]
rather than a rolling or otherwise explicitly estimated volatility measure.
This is a plausible implementation choice, but it is only one possible
interpretation of the underspecified source.

Second, the source explicitly motivates shorter quote lifetimes in volatile
periods.
The generated implementation instead uses GTC orders in both regimes and
therefore does not realize the intended quote-lifetime adaptation.

Third, the generated quote prices move inward from the displayed best bid and
best ask.
For the buy side, for example,
\[
P_t^{buy}=bid_t+\delta,
\]
which can make the order immediately marketable rather than purely passive.
This choice is potentially inconsistent with the source's stated
maker-oriented mechanism.

These discrepancies illustrate why successful code generation alone is not a
sufficient criterion for successful strategy operationalization.

\paragraph{Historical execution protocol.}
The historical evaluation for this example uses two-hour market episodes with
decisions evaluated once per minute.
Each market is assigned an independent account with an initial value of
\$1,000.
The market set is selected from the common historical LOB evaluation panel
according to data availability and coverage rather than strategy profitability.

For this particular diagnostic replay, resting maker orders are evaluated under
a relaxed execution profile that converts immediately executable portions into
aggressive historical-book matches.
Consequently, the resulting financial outcomes should be interpreted as a
\emph{sensitivity evaluation of the generated decision rule}, rather than as a
faithful measurement of passive spread capture.
In particular, the replay does not establish that the source strategy would
have achieved the same outcomes under true post-only queue dynamics. The evaluated sample contains 77 distinct market episodes spanning several
domains, as summarized in Table~\ref{app:worked_example_c029}.

\paragraph{Observed financial behavior.} Of the 77 market episodes, 52 contain at least one historical fill and are financially evaluable under the replay contract; the remainder produce no fills or fail historical data checks and are excluded rather than assigned zero returns. Across these 52 episodes the mean net ROI is $-3.59\%$ (median $-0.63\%$; 10th and 90th percentiles $\approx-13.3\%$ and $+0.05\%$), with only 6 (11.5\%) positive episodes and 1,116 fills in total. The distribution is highly heterogeneous---the strongest episode reaches $\approx+67.9\%$ ROI and the weakest $\approx-59.7\%$---but these extremes are descriptive rather than standalone evidence of profitability; and because each episode runs an independent account, episode-level ROIs are never summed into a synthetic portfolio return.

\begin{table*}[t]
\centering
\setlength{\tabcolsep}{4pt}
\renewcommand{\arraystretch}{1.05}

% ============================================================
% Panel (a)
% ============================================================
\begin{minipage}[t]{0.46\textwidth}
\vspace{0pt}
\centering
\textbf{(a) Model-specified operationalization}
\vspace{3pt}

\scriptsize
\begin{adjustbox}{max width=\linewidth}
\begin{tabular}{lll}
\toprule
\textbf{Decision}
& \textbf{High-volatility}
& \textbf{Low-volatility} \\
\midrule

Regime
& \(v_t>0.02,\;s_t<0.05\)
& \(v_t\leq0.02,\;s_t<0.02\) \\

Buy quote
& \(\max(0.01,bid_t+0.05)\)
& \(\max(0.01,bid_t+0.01)\) \\

Sell quote
& \(\min(0.99,ask_t-0.05)\)
& \(\min(0.99,ask_t-0.01)\) \\

Buy size
& \(\min(0.005\,cash/P,0.5C,5)\)
& \(\min(0.01\,cash/P,C,10)\) \\

Sell size
& \(\min(inv,0.5C,5)\)
& \(\min(inv,C,10)\) \\

Liquidity
& Maker, post-only
& Maker, post-only \\

TIF
& GTC
& GTC \\
\bottomrule
\end{tabular}
\end{adjustbox}
\end{minipage}
\hfill
% ============================================================
% Panel (b)
% ============================================================
\begin{minipage}[t]{0.20\textwidth}
\vspace{0pt}
\centering
\textbf{(b) Market domains}
\vspace{3pt}

\scriptsize
\begin{tabular}{lr}
\toprule
\textbf{Domain} & \textbf{Episodes} \\
\midrule
Sports      & 54 \\
Crypto      & 7 \\
Politics    & 6 \\
Finance     & 5 \\
Weather     & 3 \\
Technology  & 1 \\
Economics   & 1 \\
\midrule
\textbf{Total} & \textbf{77} \\
\bottomrule
\end{tabular}
\end{minipage}
\hfill
% ============================================================
% Panel (c)
% ============================================================
\begin{minipage}[t]{0.29\textwidth}
\vspace{0pt}
\centering
\textbf{(c) Financial summary}
\vspace{3pt}

\scriptsize
\begin{tabular}{lr}
\toprule
\textbf{Metric} & \textbf{Value} \\
\midrule
Market episodes              & 77 \\
Financially evaluable        & 52 \\
No-fill episodes             & 21 \\
Historical fills             & 1,116 \\
Mean net ROI                 & \(-3.59\%\) \\
Median net ROI               & \(-0.63\%\) \\
10th pct. ROI                & \(\approx-13.3\%\) \\
90th pct. ROI                & \(\approx+0.05\%\) \\
Positive episodes            & \(6/52\) \\
Positive-episode rate        & \(11.5\%\) \\
Oracle-settled               & 42 \\
Last-price-valued            & 10 \\
\bottomrule
\end{tabular}
\end{minipage}

\vspace{5pt}

\caption{
Worked example of the volatility-adaptive market-making strategy.
(a) Model-specified quantitative operationalization of the source strategy.
(b) Domain composition of the historical evaluation episodes.
(c) Financial outcomes over episodes.
}
\label{tab:c029_threeway}

\end{table*}

\begin{figure}
    \centering
    \includegraphics[width=1.0\linewidth]{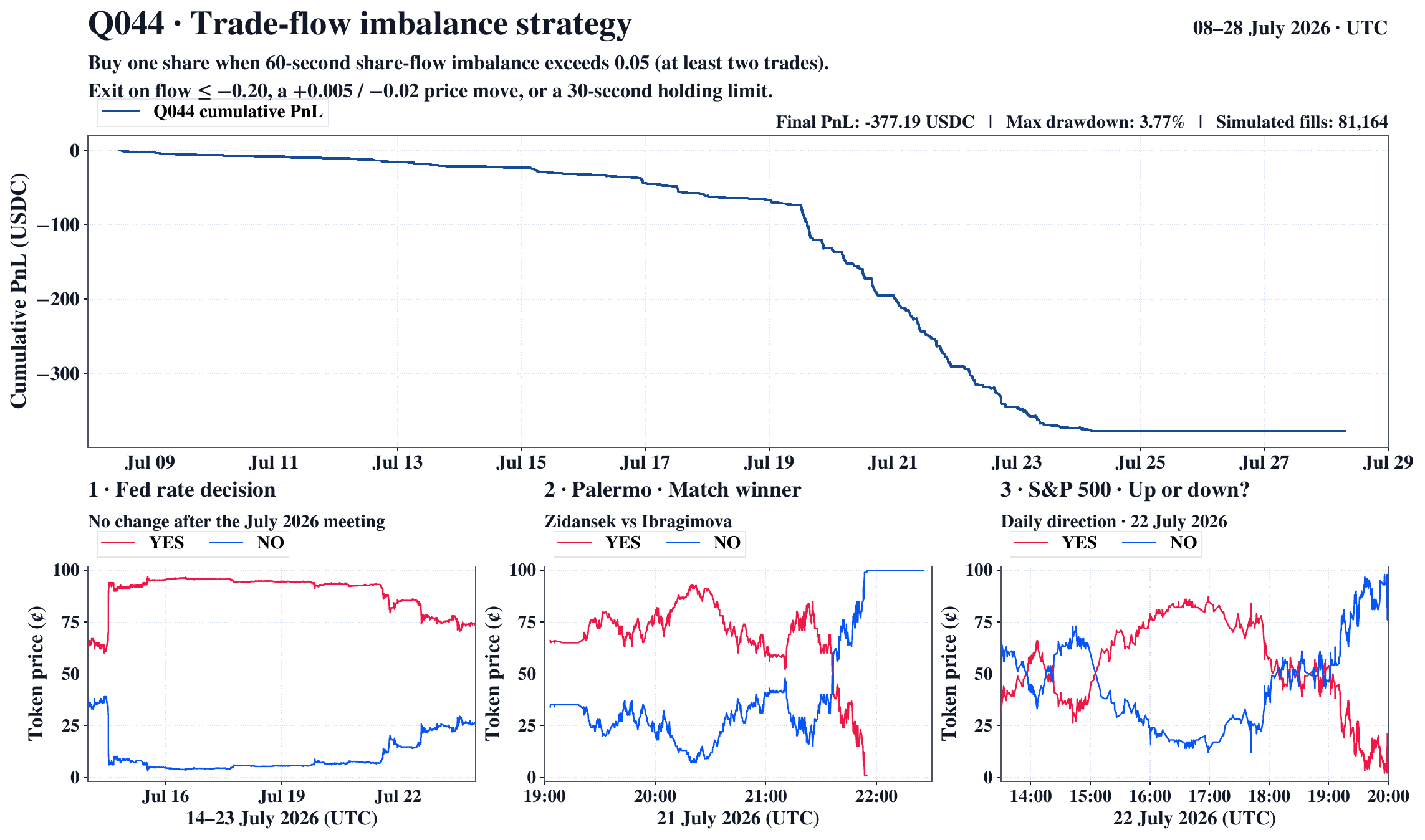}
\caption{
Historical replay of the Q044 trade-flow imbalance strategy.
\textbf{Top:} cumulative PnL from July 8--28, 2026 (UTC), with 81,164 simulated fills, final PnL of $-377.19$ USDC, and maximum drawdown of $3.77\%$.
\textbf{Bottom:} representative YES/NO price trajectories from three heterogeneous prediction markets---a Federal Reserve policy decision, a football match, and an S\&P~500 directional contract.
The examples illustrate the cross-domain heterogeneity encountered when applying a common operationalized trading rule to prediction-market episodes.
}
    \label{fig:Q044}
\end{figure}

\paragraph{Cross-domain behavior of the trade-flow imbalance strategy.}
Figure~\ref{fig:Q044} provides a representative example of how a common quantitative rule can exhibit markedly different behavior across heterogeneous prediction-market contracts. Q044 is a short-horizon trade-flow strategy that interprets recent transaction imbalance as evidence of temporary directional pressure: it buys one outcome share when the 60-second share-flow imbalance exceeds \(0.05\), subject to observing at least two trades, and exits when the flow reverses below \(-0.20\), when the token price reaches the prescribed profit/loss thresholds (\(+0.005/-0.02\)), or when the 30-second holding limit is reached. The upper panel reports the aggregate replay from July 8--28, 2026, during which the strategy generated 81,164 simulated fills but accumulated a final PnL of \(-377.19\) USDC, illustrating that frequent locally directional signals do not necessarily translate into positive aggregate alpha. The lower panels highlight three qualitatively different contracts drawn from the same evaluation universe: (1) a \emph{Federal Reserve rate-decision} market asking whether there would be no change after the July 2026 meeting; (2) a \emph{Palermo tennis match-winner} market for Zidansek versus Ibragimova; and (3) an intraday \emph{S\&P~500 Up-or-Down} market for July 22, 2026. These examples expose several prediction-market-specific features absent from conventional single-asset backtests. Prices are bounded contingent-claim values rather than unconstrained asset prices; the YES and NO tokens jointly encode beliefs about mutually exclusive outcomes and tend toward terminal payoffs as information is incorporated and resolution approaches; and the information-arrival process differs substantially across domains, ranging from gradual macroeconomic repricing, to rapidly resolving sports information, to high-frequency financial-market movements. Consequently, an identical order-flow rule is not applied to statistically interchangeable tickers: it operates over contracts with different event horizons, information processes, liquidity profiles, and resolution dynamics. The figure therefore illustrates why prediction-market strategy evaluation must account not only for realized PnL, but also for event identity, outcome semantics, temporal alignment, execution conditions, and settlement structure when assessing whether an operationalized strategy generalizes across markets.

\section{Experimental Settings}

\subsection{Evaluation Metrics}
\label{app:evaluation_metrics}

We evaluate each generated strategy from two complementary perspectives:
financial performance and prediction-market-specific executability.
The former measures whether an implemented strategy produces economically
meaningful returns under a common account and execution environment, while
the latter measures whether its signals can be supported by the historical
data and execution evidence available in prediction markets.

For \textbf{financial performance}, we report \textbf{Net ROI}, the total
return over the evaluation window after modeled trading costs;
\textbf{Settled ROI}, which replaces terminal mark-to-market value with the
verified settlement value when available; \textbf{Maximum Drawdown (MDD)},
the largest peak-to-trough decline in account equity; \textbf{Sharpe Ratio
(SR)}, which measures risk-adjusted return when a valid return series is
available; and \textbf{Cost Drag}, the reduction from gross to net PnL due
to modeled fees, slippage, and rebates. We do not annualize short-horizon
episode returns. Daily Sharpe is reported only when a complete daily NAV
series is available; otherwise the metric is marked as unavailable or
explicitly identified as a research equity-path statistic.

Prediction markets additionally require evaluating whether a strategy can
actually observe the required information and obtain executable exposure.
We therefore report \textbf{Fill Ratio}, the fraction of requested quantity
that is executed; \textbf{Input Coverage}, the fraction of decision points
for which all required causal inputs are available; \textbf{Execution
Observability}, the fraction of submitted orders for which the historical
backend contains sufficient execution evidence; \textbf{Markout}, the
signed post-fill price movement at fixed horizons; and \textbf{Settlement
Coverage}, the fraction of eligible episodes with a verified terminal
outcome. These metrics separate signal quality from data availability,
execution feasibility, and final contract resolution.

Together, the two groups distinguish profitable-looking strategies from
strategies that are both financially effective and historically executable.

\begin{table*}[t]
\centering
\caption{
Evaluation metrics used in our benchmark.
$V_t$ denotes account equity at time $t$, $V_0$ the initial capital,
and $s_j\in\{+1,-1\}$ the BUY/SELL direction.
$\uparrow$ ($\downarrow$) indicates higher (lower) is better.
}
\label{tab:evaluation_metrics}
\small
\setlength{\tabcolsep}{4pt}
\renewcommand{\arraystretch}{1.15}

\begin{tabularx}{\textwidth}{
    l
    c
    >{\raggedright\arraybackslash}p{0.30\textwidth}
    >{\raggedright\arraybackslash}X
}
\toprule
\textbf{Metric} & \textbf{Dir.} & \textbf{Formula} & \textbf{Description} \\
\midrule

\multicolumn{4}{l}{\textbf{E3: Financial performance}} \\

Net ROI
& $\uparrow$
& $\displaystyle
\mathrm{ROI}
=
\frac{V_T-V_0}{V_0}$
& Total return over the evaluation window; not annualized. \\

Settled ROI
& $\uparrow$
& $\displaystyle
\mathrm{ROI}_{\mathrm{settled}}
=
\frac{V_T^{\mathrm{settled}}-V_0}{V_0}$
& Return after terminal settlement; research-price fallback is reported separately. \\

MDD
& $\downarrow$
& $\displaystyle
\max_t
\frac{\max_{u\le t}V_u-V_t}
     {\max_{u\le t}V_u}$
& Maximum peak-to-trough decline of account equity. \\

Cost Drag
& $\downarrow$
& $\displaystyle
\mathrm{CD}
=
\mathrm{PnL}_{gross}
-
\mathrm{PnL}_{net}$
& PnL lost to modeled fees, slippage, and other execution costs net of rebates. \\

\midrule
\multicolumn{4}{l}{\textbf{E4: Prediction-market-specific evaluation}} \\

Fill Ratio
& $\uparrow$
& $\displaystyle
\frac{\sum_o q_o^{fill}}
     {\sum_o q_o^{req}}$
& Fraction of requested quantity that is executed. \\

Input Coverage
& $\uparrow$
& $\displaystyle
\frac{N_{\mathrm{valid\ input}}}
     {N_{\mathrm{decision}}}$
& Fraction of decision points with all required causal inputs available. \\

Execution Observability
& $\uparrow$
& $\displaystyle
\frac{N_{\mathrm{observable\ orders}}}
     {N_{\mathrm{submitted\ orders}}}$
& Fraction of submitted orders for which the historical backend provides execution evidence. \\

Markout$_{\Delta}$
& $\uparrow$
& $\displaystyle
\frac{1}{N}
\sum_j
s_j
\left(
p_{t_j+\Delta}-p_j^{exec}
\right)$
& Signed post-fill price movement, measured at the next decision point and at 5 minutes. \\

Settlement Coverage
& $\uparrow$
& $\displaystyle
\frac{N_{\mathrm{verified\ settlement}}}
     {N_{\mathrm{settlement\ required}}}$
& Fraction of eligible episodes with verified terminal resolution. \\

\bottomrule
\end{tabularx}
\end{table*}

\subsection{Experimental Cohorts and Aggregation}
\label{app:experimental_cohorts}

The experiments in AlphaOpsBench use several distinct populations that
serve different analytical purposes. We retain separate identities for
the main generation cohort, research-executable replay packages, legacy
diagnostics, independent full-chain repetitions, execution interventions,
and engineering benchmarks. These populations are not pooled merely
because they share task identifiers.

The main benchmark contains 477 active task identities and 954 planned
task--method slots. The historical research-replay population contains
456 method-specific packages covering 390 distinct tasks; this population
is not equivalent to the set of original candidates passing all semantic
checks. The latest independent-generation campaign contains 36 controlled
tasks and 24 preregistered real-strategy tasks, evaluated under Direct and
Staged generation with five independent repetitions at $T=0.7$ and $K=1$,
yielding 600 full-chain generation attempts. Older shared-Designer $K=5$
runs and the legacy $T=0$ path analysis are retained as separate
diagnostics.

Coverage statistics use their declared planned denominators.
Financial aggregation first removes duplicate execution identities within
task and method, aggregates eligible episodes within each task, and then
weights tasks equally. Each metric retains its own eligibility denominator.
Main financial uncertainty intervals use task-block bootstrap resampling
with 2,000 resamples; these intervals preserve task-level aggregation but
do not eliminate all dependence arising from shared markets or events.
Missing valuations, infrastructure failures, unevaluated stages, and
evaluated failures are retained as distinct states.

For supplementary raw-policy analyses, we distinguish full-account
performance from exposure-conditioned performance. A valid strategy that
executes with all required inputs, encounters no candidate error, maintains
a complete account and valuation path, and legitimately takes no position
may have an observed zero ROI and zero drawdown. In contrast, an
input-guarded, unevaluated, or failed execution is not assigned a synthetic
zero return. This convention is reported separately and does not
retroactively change the frozen eligibility definition used by earlier
experiments.

\begin{table*}[t]
    \centering
    \small
    \caption{
    Experimental cohorts and aggregation units.
    Planned generation slots, completed replay units, and financially
    evaluable observations are different denominators.
    }
    \label{tab:experimental_cohorts}
    \resizebox{\textwidth}{!}{
    \begin{tabular}{lllll}
        \toprule
        Track & Population & Protocol & Recorded extent & Statistical unit \\
        \midrule
        Main generation
        & 477 active tasks
        & $T=0.7$, $R=1$, $K=1$
        & 954 planned slots; 952 terminals
        & task--method \\

        Research B1
        & 456 packages; 390 task identities
        & separately routed replay cohorts
        & 783,655 planned; 775,725 COMPLETE
        & deduplicated episode, then task \\

        Legacy paths
        & 12 diagnostic task identities
        & $T=0$ historical diagnostic
        & unequal episode support
        & episode distribution \\

        Independent R5
        & 36 controlled + 24 real tasks
        & $T=0.7$, $R=5$, $K=1$
        & 600 full-chain attempts
        & task and repetition \\

        Native execution sensitivity
        & 15 tasks; 19 task--market windows
        & fee / size / latency / capital / capacity
        & 529 configurations
        & within-window change \\

        Frozen fee sweep
        & 59,356 PML2 + 153,907 V3 episodes
        & alternative fee assumptions
        & 213,263 episodes
        & episode; no new strategy decision \\

        Legacy raw / overlay
        & 159 submitted jobs per arm
        & older research-input assumptions
        & 157 common completed pairs
        & matched candidate--window \\

        Efficiency
        & frozen finite workloads
        & three physical repetitions
        & 63 comparable pairs
        & within-engine workload \\
        \bottomrule
    \end{tabular}}
\end{table*}

\section{Detailed Results}

\subsection{Historical Strategy Paths by Specification Level}
\label{app:strategy_paths}

\begin{figure}
    \centering
    \includegraphics[width=1\linewidth]{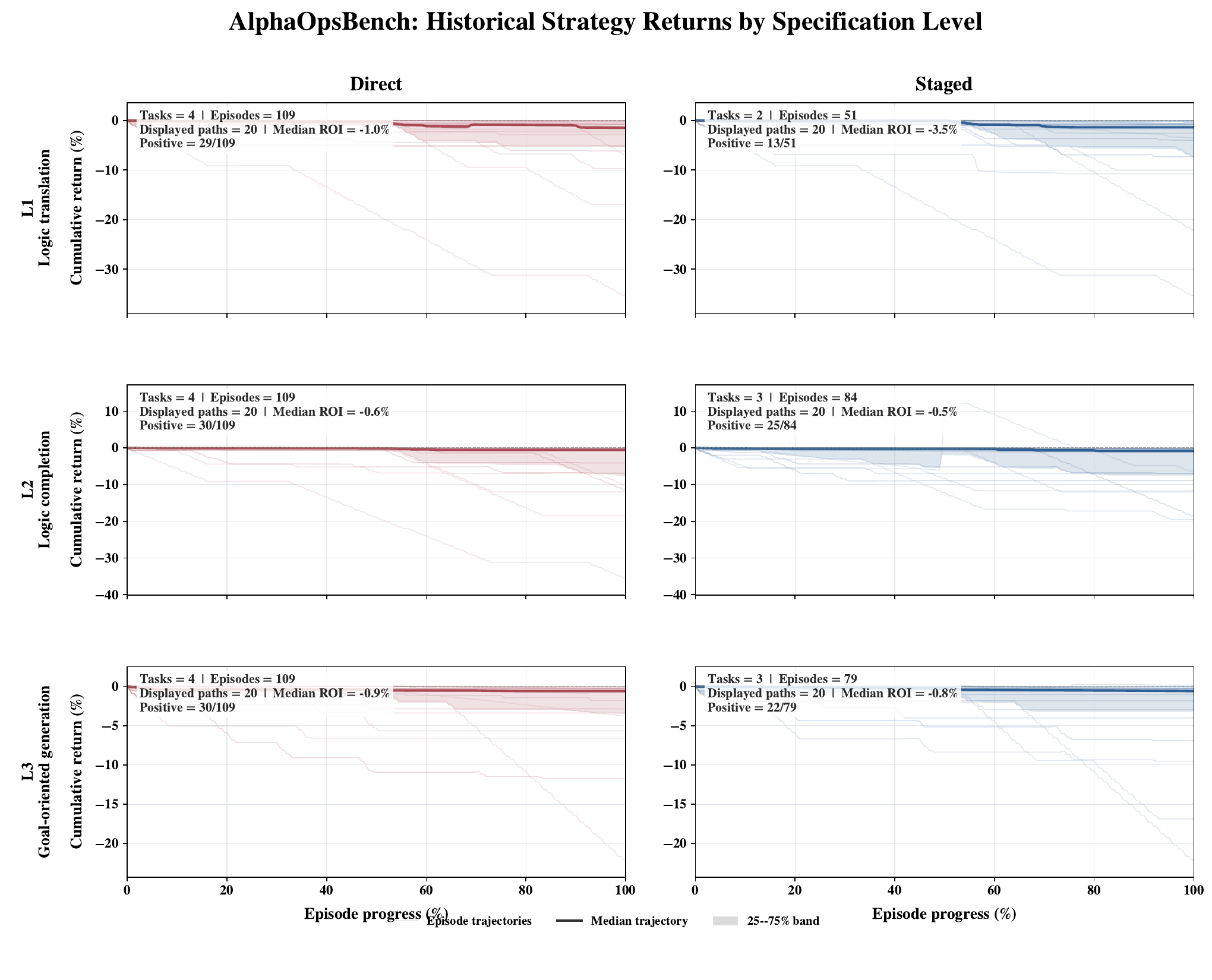}
    \caption{Historical strategy-return trajectories across specification levels under Direct and Staged operationalization.}
    \label{fig:strategy_paths}
\end{figure}

Figure~\ref{fig:strategy_paths} reports episode-level cumulative returns of PML2 replay episodes generated at temperature 0.0---the main protocol samples at 0.7, so this is a descriptive appendix view---with one panel per specification level and arm. The six cohorts are unpaired and unequal (Direct: four tasks and 109 eligible episodes per level; Staged: two, three, and three tasks with 51, 84, and 79 episodes), and each path is an independent episode aligned by normalized episode progress rather than a shared portfolio series. Returns concentrate near zero with a downside tail of roughly 20--35 percentage points: medians lie between $-0.5\%$ and $-1.0\%$ in five panels, with L1 Staged lowest at $-3.5\%$ on only two tasks, and positive-episode shares stay within 25--30\% everywhere ($29/109$ vs.\ $13/51$, $30/109$ vs.\ $25/84$, $30/109$ vs.\ $22/79$). Neither arm orders monotonically across L1--L3, and within-panel dispersion exceeds between-arm and between-level differences, so the figure characterizes return distributions under the stated execution assumptions rather than supporting causal method comparisons.

\subsection{Independent-Generation Stability}
\label{app:independent_generation_stability}

\paragraph{Protocol.}
We evaluate run-to-run reproducibility using independent full-chain
generation with $R=5$, $K=1$, and $T=0.7$. The controlled cohort contains
36 tasks spanning six mechanisms, two base tasks per mechanism, and three
specification levels. A separate source-derived cohort contains 24
preregistered real strategy tasks. Both Direct and Staged are regenerated
independently in every repetition; in particular, every Staged repetition
produces a new Designer output before the Coder stage. Invalid upstream
designs that prevent coding remain observed failures of that generation
chain rather than being resampled.

This protocol yields 600 completed full-chain attempts:
360 controlled attempts and 240 Real24 attempts. Saved responses are reused
only when their generation identities match exactly, while evaluator
projection is performed offline. The controlled and Real24 cohorts are
reported separately.

\paragraph{Canonical validity across repetitions.}
Controlled Direct and Staged obtain 35/180 and 20/180 canonical passes,
respectively. At L1, neither method obtains a canonical pass; at L2 the
counts are 17/60 and 9/60, and at L3 they are 18/60 and 11/60.
The Real24 cohort has no confirmed canonical pass under either method.
Because the evidence requirements and evaluability of individual checks
differ across specification levels, these values should not be interpreted
as a difficulty ranking.

We distinguish four repetition statistics. First-sample Pass@1 uses only
repeat 1. Attempt success averages all five independent attempts.
Pass@5 asks whether at least one of the five repetitions passes, whereas
``all five'' requires canonical acceptance in every repetition.

\begin{table}[t]
    \centering
    \small
    \caption{
    Canonical success under independent full-chain generation.
    First-sample Pass@1 uses repeat 1; attempt success pools all five
    repetitions. UNVERIFIED counts attempts rather than tasks.
    }
    \label{tab:r5_success}
    \begin{tabular}{llccccc}
        \toprule
        Cohort & Method & Pass@1 & Attempt success & Pass@5 & All five & Unverified \\
        \midrule
        Controlled & Direct & 4/36 & 35/180 & 13/36 & 2/36 & 15 \\
        Controlled & Staged & 3/36 & 20/180 & 10/36 & 1/36 & 22 \\
        Real24 & Direct & 0/24 & 0/120 & 0/24 & 0/24 & 17 \\
        Real24 & Staged & 0/24 & 0/120 & 0/24 & 0/24 & 12 \\
        \bottomrule
    \end{tabular}
\end{table}

\paragraph{Preservation and completion of economic decisions.}
For controlled tasks, source-specified and model-owned decisions are
evaluated separately. L1 primarily tests preservation of source-given
economic choices. L2 combines binding source requirements with model-owned
parameter completion, whereas L3 primarily evaluates model-owned
operationalization. Conditional preservation or completion rates are always
reported together with their evidence coverage; unverified slots are not
converted into failures or passes.

\begin{table}[t]
    \centering
    \small
    \caption{
    Controlled operationalization by specification level.
    Given/open columns operate on economic decision slots, whereas canonical
    columns operate on generation attempts. Conditional rates must be read
    together with their evidence coverage.
    }
    \label{tab:controlled_operationalization}
    \begin{tabular}{llcccc}
        \toprule
        Level & Method & Canonical pass & Canonical cov. &
        Given preservation / cov. & Open completion / cov. \\
        \midrule
        L1 & Direct & 0/60 & 75.0\% & 93.8\% / 11.9\% & NA \\
        L1 & Staged & 0/60 & 78.3\% & 94.7\% / 7.0\% & NA \\
        L2 & Direct & 17/60 & 98.3\% & 82.9\% / 17.5\% & 44.1\% / 89.4\% \\
        L2 & Staged & 9/60 & 96.7\% & 90.0\% / 10.0\% & 23.5\% / 95.6\% \\
        L3 & Direct & 18/60 & 100.0\% & NA & 41.5\% / 91.5\% \\
        L3 & Staged & 11/60 & 86.7\% & NA & 27.6\% / 92.6\% \\
        \bottomrule
    \end{tabular}
\end{table}

\paragraph{Design consistency.}
We compare independently generated designs through nine normalized economic
slots: data, factor, window, threshold, direction, order policy, sizing,
exit, and no-trade behavior. Agreement is computed within task over
observable repetition pairs and then aggregated at task level.
Missing designs are neither agreements nor disagreements; observed-pair
coverage is therefore reported alongside conditional agreement.

\begin{figure*}[t]
    \centering
    \includegraphics[width=\textwidth]
    {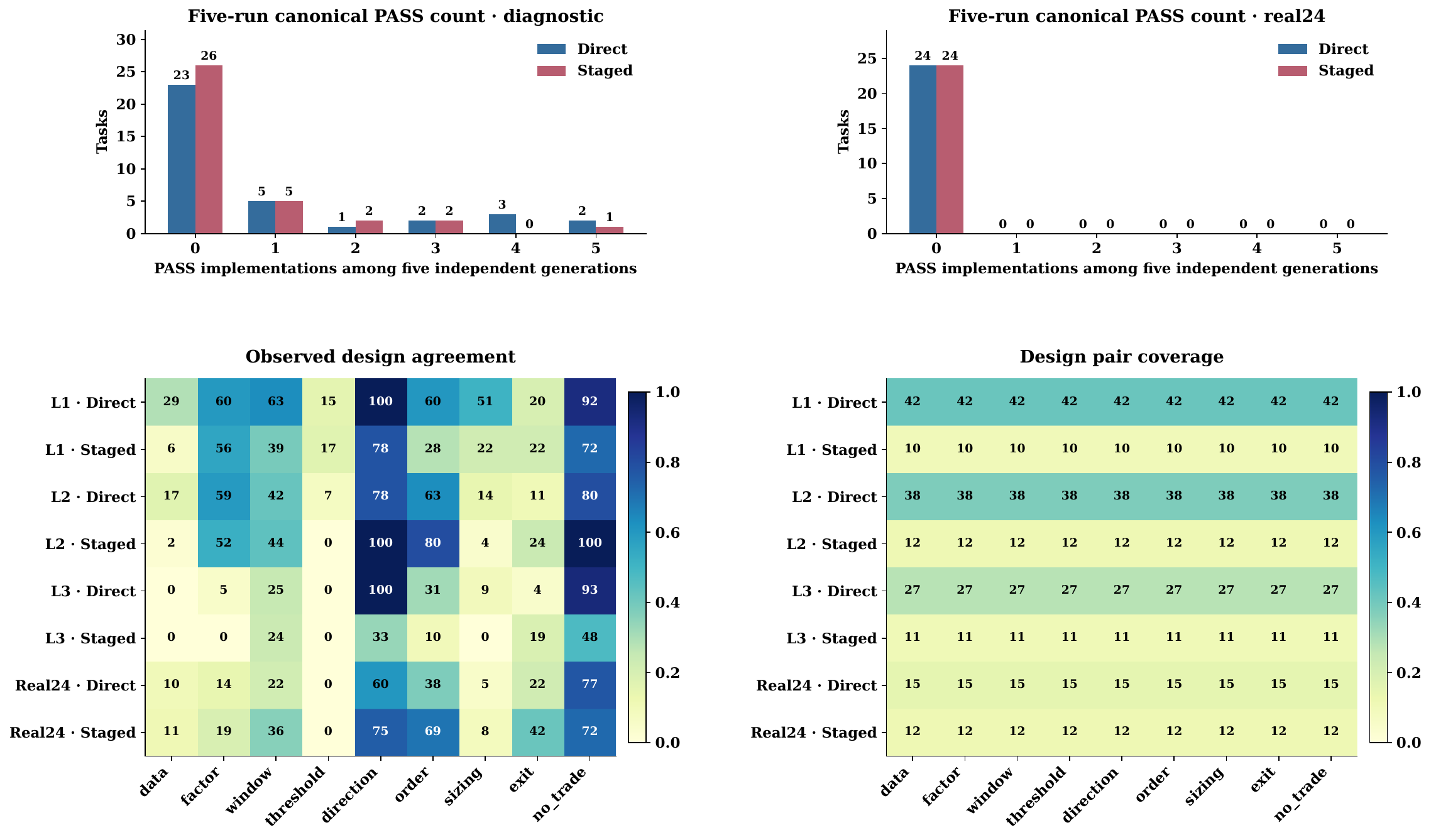}
    \caption{
    Independent-generation validity and design reproducibility over
    600 full-chain attempts. Top: number of canonical passes obtained
    across five repetitions for each task. Bottom: agreement across nine
    normalized economic design slots and the corresponding fraction of
    planned repetition pairs that are jointly observable. Conditional
    agreement should be interpreted together with pair coverage.
    }
    \label{fig:r5_generation_design}
\end{figure*}

\paragraph{Common-state behavioral consistency.}
To separate variation in generated policy logic from variation in historical
market trajectories, repeated candidates are evaluated on
candidate-independent constructed states with identical market and account
inputs. We report action agreement over all common states and separately
over reference-active states, together with the fraction of planned
candidate pairs that produce jointly observable outputs.

For L1, observed actions can additionally be compared with explicit
reference-action constraints. For L2 and L3, the diagnostic checks universal
safety constraints and consistency with each candidate's declared policy.
These checks are distinct from source fidelity. A HOLD--HOLD match
contributes to behavioral agreement but is not, by itself, evidence that
the source strategy has been faithfully implemented.

\begin{figure*}[t]
    \centering
    \includegraphics[width=\textwidth]
    {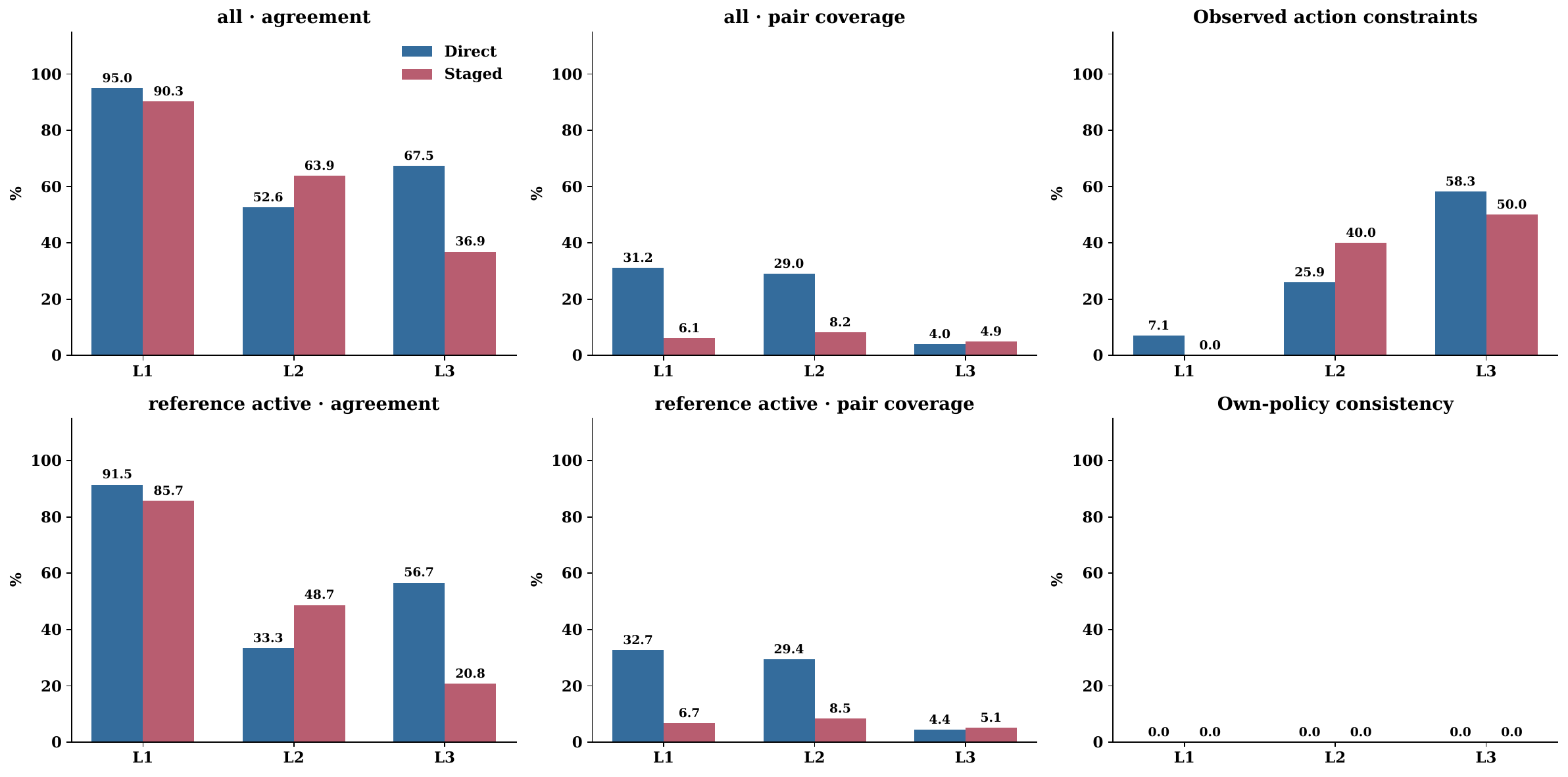}
    \caption{
    Common-state behavioral reproducibility. Agreement and pair coverage
    are task-equal and are reported for all common states and
    reference-active states. Correctness panels are conditional on observed
    candidate outputs and do not replace source-fidelity evaluation.
    }
    \label{fig:r5_common_state}
\end{figure*}

\paragraph{Raw-policy historical replay coverage.}
The frozen raw-policy experiment contains 960 planned candidate--window
slots. Of these, 495 are classified as \textsc{Input-Unavailable},
437 as \textsc{Policy-Not-Runnable}, 24 as valid HOLD, one as
\textsc{Signal-Not-Triggered}, and three as
\textsc{Order-Submitted-No-Fill}. These outcomes are retained as distinct
states. Legacy research-overlay financial outcomes are not substituted for
the missing raw-policy evidence in this campaign.

\subsection{Main-Cohort Coverage and Historical Performance}
\label{app:main_cohort_results}

The main historical evaluation is conducted on the subset of generated
candidates for which a research-executable artifact and an applicable
historical replay route are available. This population is distinct from
the canonical generation-acceptance population discussed above:
availability of a research replay package does not imply that the original
one-shot generation satisfies all source-fidelity and behavioral checks.
We therefore report replay coverage and financial outcomes separately.

\paragraph{Replay coverage.}
Table~\ref{tab:main_replay_coverage} summarizes the four routed
engine--method cohorts. PML2 contains 381 Direct and 75 Staged candidate
packages, whereas the V3 route contains 324 Direct and 10 Staged packages.
Across the four cohorts, the experiment schedules 783,655 replay slots,
of which 775,725 terminate with a \textsc{Complete} engine state,
7,928 fail, and two PML2-Staged slots terminate as partial runs.

Planned replay slots are not equivalent to independent physical executions.
In particular, the V3-Direct schedule contains 321,400 planned slots but
320,419 unique execution identities because some scheduled references map
to the same physical replay. Financial statistics are therefore
deduplicated by execution identity before aggregation. The same distinction
is retained for financial records: V3-Direct contains 306,020 financially
evaluable records before identity deduplication and 305,969 unique
financial executions.

\begin{table*}[t]
    \centering
    \small
    \setlength{\tabcolsep}{4.3pt}
    \caption{
    Main-cohort historical replay coverage.
    Planned slots denote scheduled replay units, while unique executions
    deduplicate repeated references to the same physical execution.
    \textsc{No fills} is retained as an execution outcome and is not
    converted into a zero financial return.
    }
    \label{tab:main_replay_coverage}
    \begin{tabular}{llrrrrrrrr}
        \toprule
        Engine & Method &
        Routed tasks &
        Planned &
        Unique exec. &
        Complete &
        Failed &
        Partial &
        No fills &
        Unique financial \\
        \midrule
        PML2 & Direct
        & 381 & 377,255 & 377,255 & 377,255 & 0 & 0
        & 91,084 & 281,139 \\

        PML2 & Staged
        & 75 & 75,000 & 75,000 & 74,998 & 0 & 2
        & 18,969 & 55,349 \\

        V3 & Direct
        & 324 & 321,400 & 320,419 & 313,657 & 7,743 & 0
        & 7,637 & 305,969 \\

        V3 & Staged
        & 10 & 10,000 & 10,000 & 9,815 & 185 & 0
        & 68 & 9,747 \\
        \bottomrule
    \end{tabular}
\end{table*}

\paragraph{Financial eligibility and aggregation.}
Financial analysis is restricted to unique executions with sufficient
execution and terminal-valuation evidence. A completed replay is therefore
not necessarily financially evaluable. In particular, a replay may
complete without obtaining executable exposure, or may lack the evidence
required for the declared terminal valuation. Such outcomes remain in the
coverage accounting but are not assigned a synthetic zero return.

Because individual tasks can generate very different numbers of market
episodes, we use task-equal rather than episode-equal aggregation. Let
$\mathcal{E}_q$ denote the financially evaluable executions for task $q$
and $r_{qi}$ the Net ROI of execution $i$. We first compute the mean return
within each task,
\[
    \bar r_q
    =
    \frac{1}{|\mathcal{E}_q|}
    \sum_{i\in\mathcal{E}_q} r_{qi},
\]
and then aggregate across the set $\mathcal{Q}$ of financially evaluable
tasks,
\[
    \bar r_{\mathrm{task}}
    =
    \frac{1}{|\mathcal{Q}|}
    \sum_{q\in\mathcal{Q}} \bar r_q.
\]
The reported median, quantiles, empirical distribution, and positive-task
rate are likewise computed over task-level mean returns rather than over
the pooled episode population. Task-block uncertainty intervals preserve
the task as the resampling unit.

\begin{table*}[t]
    \centering
    \small
    \setlength{\tabcolsep}{5.0pt}
    \caption{
    Task-equal historical financial outcomes for research-executable
    candidates. ``Financial exec.'' counts unique financially evaluable
    execution identities. Q10, median, and Q90 are quantiles of the
    within-task mean Net ROI distribution. The task-block interval is
    computed with tasks as the resampling unit. Positive denotes the
    fraction of tasks with positive within-task mean Net ROI.
    }
    \label{tab:main_financial_results}
    \begin{tabular}{llrrrrrrr}
        \toprule
        Engine & Method &
        Tasks &
        Financial exec. &
        Mean ROI &
        Q10 &
        Median &
        Q90 &
        Positive \\
        \midrule
        PML2 & Direct
        & 380 & 281,139
        & $-3.476\%$
        & $-3.555\%$
        & $-3.450\%$
        & $-3.399\%$
        & $0.00\%$ \\

        PML2 & Staged
        & 75 & 55,349
        & $-3.440\%$
        & $-3.516\%$
        & $-3.421\%$
        & $-3.354\%$
        & $0.00\%$ \\

        V3 & Direct
        & 323 & 305,969
        & $+0.836\%$
        & $+0.543\%$
        & $+0.917\%$
        & $+1.007\%$
        & $99.69\%$ \\

        V3 & Staged
        & 10 & 9,747
        & $+0.646\%$
        & $+0.347\%$
        & $+0.669\%$
        & $+0.885\%$
        & $100.00\%$ \\
        \bottomrule
    \end{tabular}
\end{table*}

For completeness, the task-block intervals around the task-equal mean ROI
are $[-3.515\%,-3.436\%]$ for PML2--Direct,
$[-3.504\%,-3.386\%]$ for PML2--Staged,
$[+0.814\%,+0.857\%]$ for V3--Direct, and
$[+0.496\%,+0.788\%]$ for V3--Staged.
These intervals quantify uncertainty across tasks within each routed
population; they do not make the four populations directly comparable.

\begin{figure*}[t]
    \centering
    \includegraphics[width=\textwidth]
    {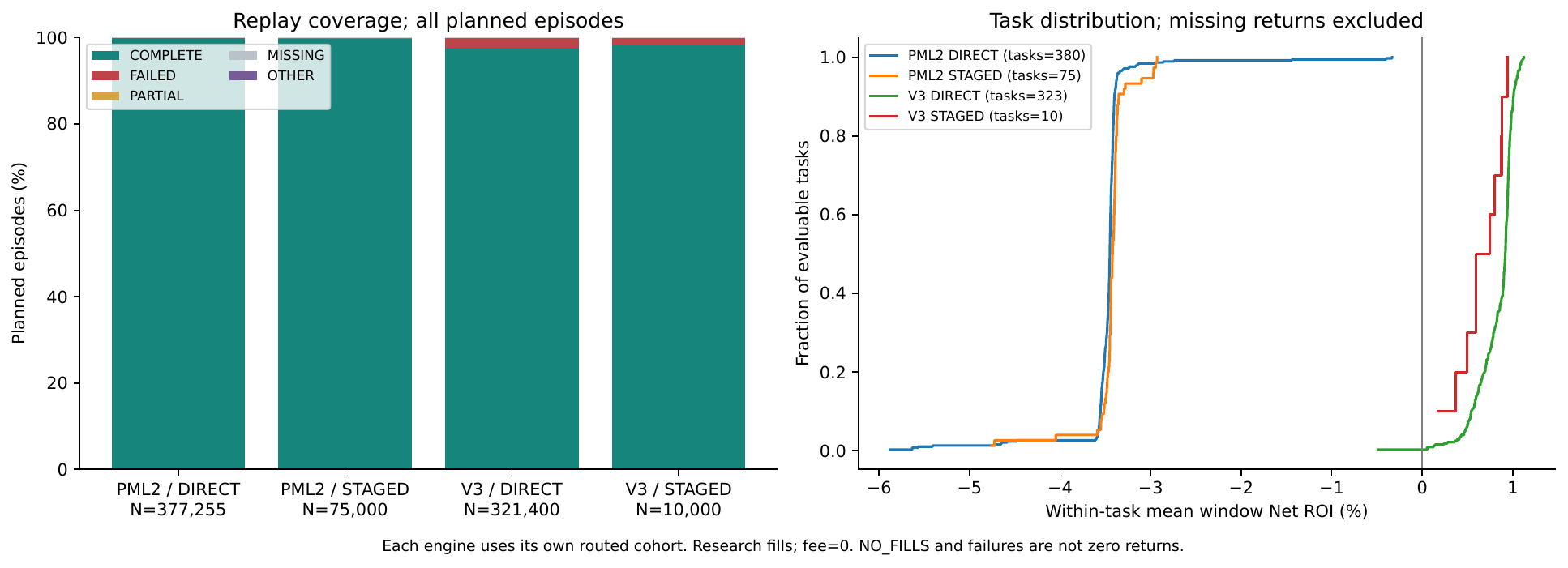}
    \caption{
    Main-cohort historical replay coverage and task-level return
    distributions.
    \textbf{Left:} outcome composition over all planned replay slots for
    each engine--method cohort.
    \textbf{Right:} empirical distributions of within-task mean window
    Net ROI among financially evaluable tasks; every task contributes one
    observation irrespective of its number of replay episodes.
    Missing financial outcomes, failed replays, and zero-fill episodes are
    retained in their corresponding coverage states rather than converted
    into zero returns.
    PML2 and V3 use independently routed candidate populations and
    different execution assumptions; the displayed return distributions
    therefore characterize each declared replay setting rather than a
    paired cross-engine performance comparison.
    }
    \label{fig:main_replay_results}
\end{figure*}

\paragraph{Observed return distributions.}
The two replay settings produce sharply different task-level return
distributions. Under PML2, the task-equal mean Net ROI is $-3.476\%$ for
Direct and $-3.440\%$ for Staged, with medians of $-3.450\%$ and
$-3.421\%$, respectively. No financially evaluable PML2 task has a
positive within-task mean return under these research settings. In V3,
the corresponding task-equal means are $+0.836\%$ for Direct and
$+0.646\%$ for Staged, and nearly all financially evaluable tasks have a
positive within-task mean return.

This sign separation should not be interpreted as evidence that V3 is
financially superior to PML2. The engines operate on different routed
candidate populations and implement materially different execution
assumptions. V3 provides broad trade-driven replay under modeled
executable liquidity, whereas PML2 conditions execution on reconstructed
historical order-book state and associated readiness constraints.
Differences in candidate support, execution observability, fill formation,
window construction, and terminal valuation can therefore contribute to
the observed separation.

\paragraph{Interpretation.}
These results establish the scale and financial coverage of the main
historical evaluation rather than a profitability claim. The benchmark
contains hundreds of research-executable strategy tasks and hundreds of
thousands of unique financially evaluable historical executions, but the
observed financial outcomes depend on the evidence and execution contract
under which each candidate is evaluated. Reporting replay coverage,
financial eligibility, and task-equal returns jointly prevents unavailable
execution evidence or zero-fill behavior from being silently absorbed into
the return distribution, and separates the question of whether a strategy
can be evaluated from the question of how that strategy performs once
financially evaluable.

\begin{figure*}[t]
    \centering
    \includegraphics[width=\textwidth]{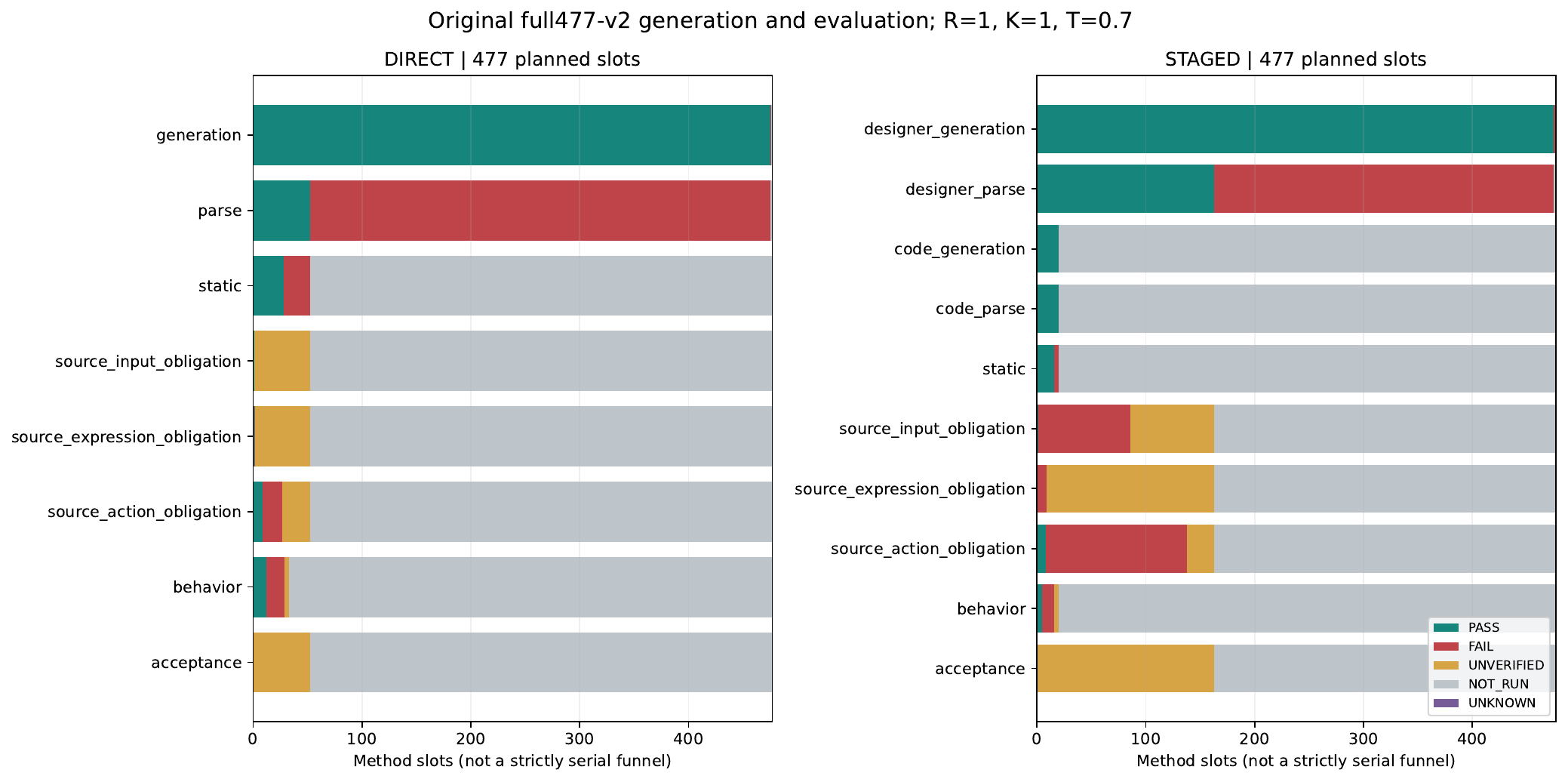}
    \caption{
    Stage-wise generation and evaluation outcomes for Direct and Staged
    on the main benchmark. Each row accounts for 477 planned task--method
    slots, distinguishing passes, failures, unverified outcomes,
    unexecuted stages, and unknown states. Rows are not successive
    survival counts in a strictly serial funnel.
    }
    \label{fig:generation_outcomes}
\end{figure*}

\paragraph{Generation outcomes and failure attribution.}
Figure~\ref{fig:generation_outcomes} reports stage-wise outcomes for the
477-task main cohort under $R=1$, $K=1$, and $T=0.7$.
Generation is attempted for 476 tasks per method, with the excluded task
retained in the coverage denominator. Initial responses are available for
nearly all tasks, but parsing rejects a substantial fraction of outputs.
Staged additionally shows limited progression from design parsing to code
generation. Subsequent source-requirement and behavioral checks contain
both explicit failures and cases without a conclusive evaluation.
The rows therefore describe stage-specific evidence rather than a strictly
serial funnel; unverified, unexecuted, and unknown outcomes are distinguished
from confirmed failures. Neither method records a canonical
\textsc{CandidatePass} in the original evaluation.
Separately, 381 Direct and 75 Staged candidates are available for research
replay, including recovered exports and research execution variants.
These counts measure research executability, not acceptance of the original
generated implementations.

\begin{figure}
    \centering
    \includegraphics[width=1\linewidth]{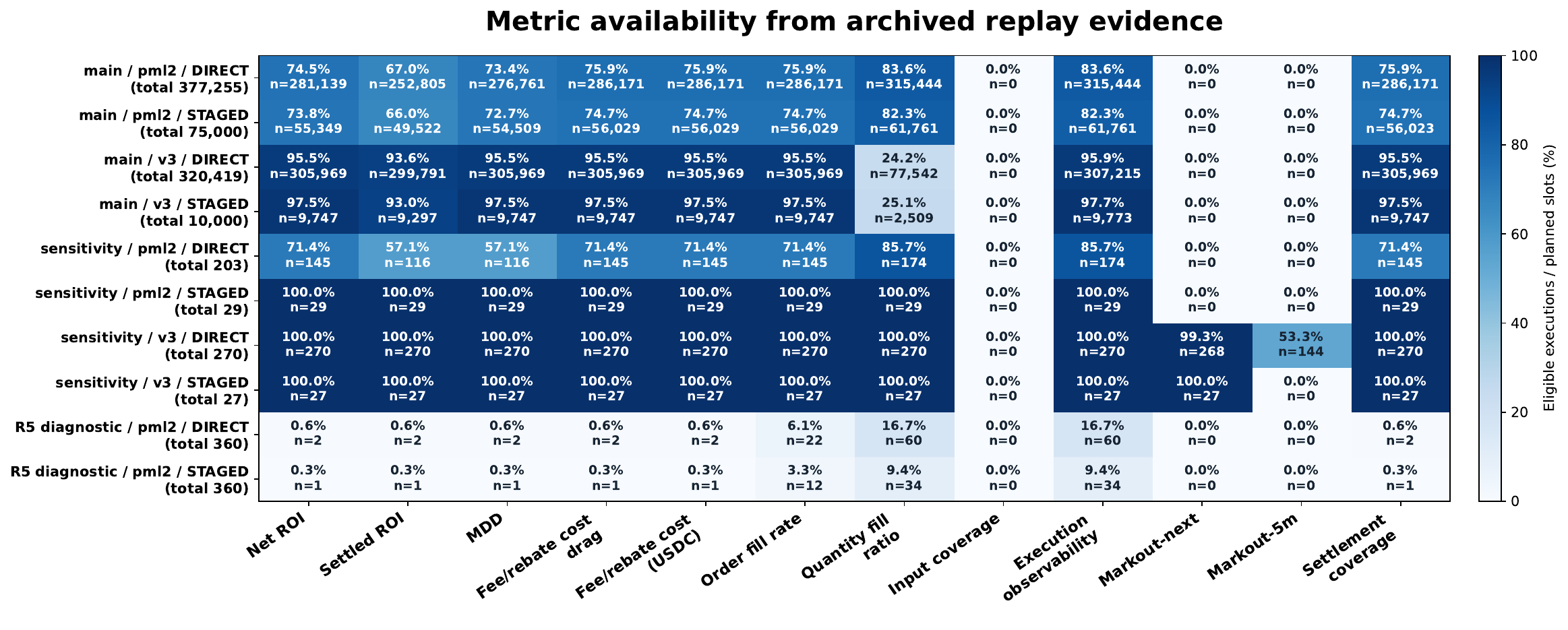}
    \caption{Metric Availability}
    \label{fig:metric_availability}
\end{figure}
\subsection{Metric Availability and Evidence Coverage}
\label{app:metric_availability}

Metric availability is itself part of the benchmark outcome because different
statistics require different historical evidence. Figure~\ref{fig:metric_availability}
reports the fraction of planned replay units for which each metric can be
computed from archived evidence. Main-cohort financial metrics have broad
coverage, but microstructure-specific statistics are substantially more
selective. For example, V3 supports Net ROI on more than \(95\%\) of its main
replay slots, whereas quantity-fill ratios are available for only about one
quarter of the same population. Input-coverage and markout statistics cannot
be reconstructed for the main archived cohorts and are therefore reported as
unavailable rather than zero.

The pattern also differs across experimental tracks. Controlled execution
sensitivity runs preserve richer execution evidence and consequently support
markout measurements in subsets where the main replay does not. In contrast,
the independent-generation R5 experiments have much lower financial evidence
coverage and are used primarily to study generation and behavioral
reproducibility. These differences motivate the metric-specific denominators
reported throughout the appendix and prevent unavailable evidence from being
silently converted into favorable or unfavorable performance.

\section{Additional Findings}
\label{app:findings}

This section complements the main benchmark evaluation with execution-cost
diagnostics. We distinguish two forms of evidence. First, population-scale
counterfactual analyses apply alternative fee or reference-price assumptions
to previously recorded execution paths. Second, controlled paired replays
rerun the strategy through the native account and execution logic, allowing
fees, capital, size, and latency to affect subsequent trading decisions.
The former provides broad sensitivity estimates, while the latter tests when
a frozen-path approximation remains valid.

% ============================================================
\subsection{Transaction Fees in Polymarket}
\label{app:transaction_fees}
% ============================================================

\paragraph{Population-scale fee sensitivity.}
We first evaluate transaction costs on frozen execution paths, holding signals,
orders, fills, execution prices, and terminal valuation fixed. The population
contains \(59{,}356\) PML2 and \(153{,}907\) Fill-only V3 episodes.
For a nonlinear fee schedule, a fill of \(C\) shares at probability price
\(p\) incurs
\begin{equation}
    f = C\,r\,p(1-p).
    \label{eq:nonlinear_fee}
\end{equation}

Under the category-policy counterfactual used in the original analysis, mean
ROI decreases by \(0.091\) percentage points in PML2 and \(0.719\) percentage
points in V3. Only \(5/16{,}214\) initially positive PML2 episodes cross zero,
compared with \(7{,}799/82{,}200\) (\(9.49\%\)) in V3. A broader fee sweep
shows the same qualitative pattern: narrow positive margins are more sensitive
to transaction costs than already negative episodes. These figures describe
fee sensitivity under the recorded execution paths rather than reconstructed
historical Polymarket fee bills.

\begin{figure}[t]
    \centering
    \includegraphics[width=\linewidth]{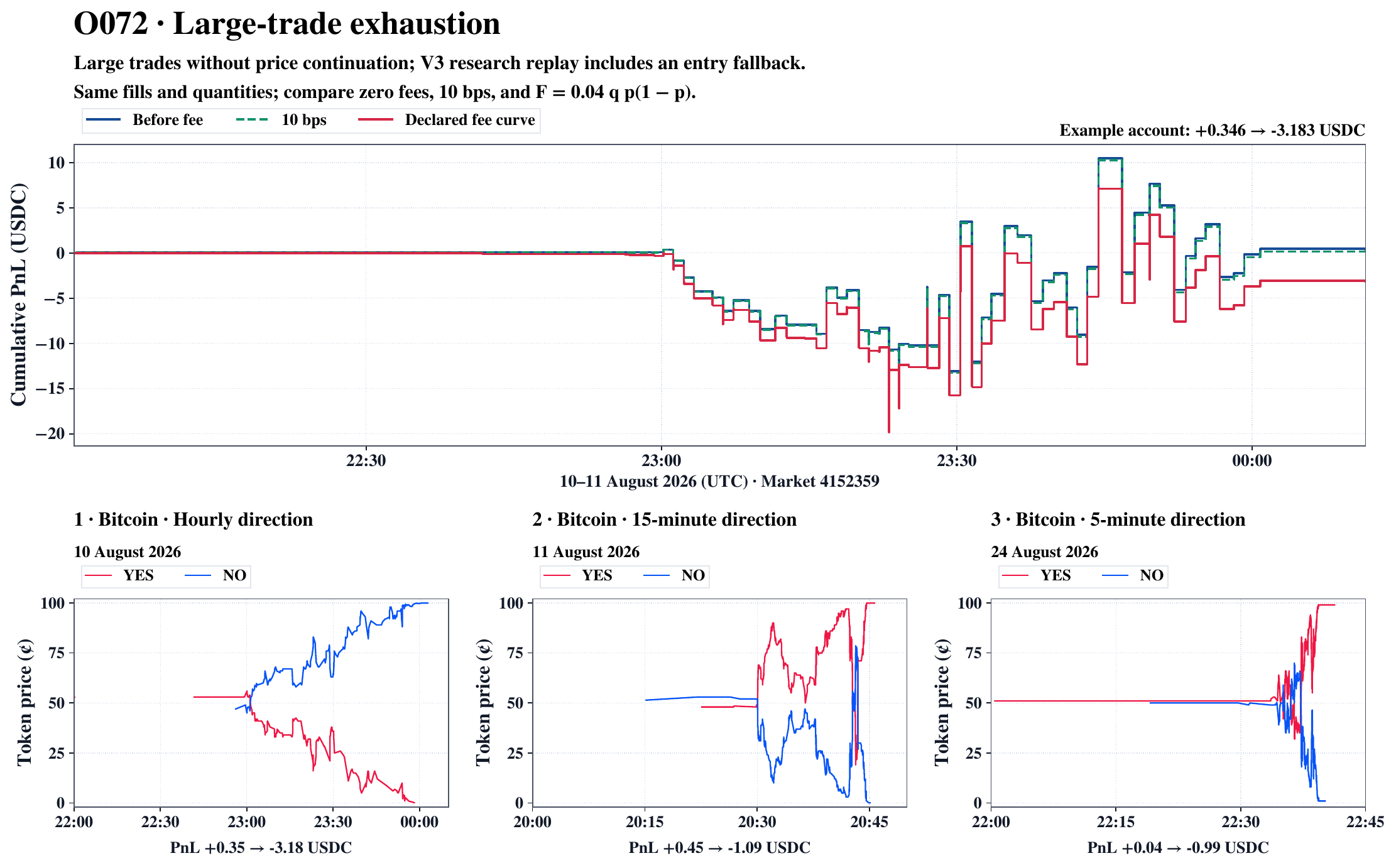}
    \caption{
    Fee sensitivity of the Q072 large-trade exhaustion strategy under fixed
    execution paths. The same signals, fills, and quantities are evaluated
    under alternative fee assumptions; representative short-horizon Bitcoin
    contracts illustrate how a small pre-fee edge can become negative after
    costs.
    }
    \label{fig:fee_sensitivity}
\end{figure}

% ============================================================
\subsection{Execution-Price Deviation and Depth}
\label{app:slippage}
% ============================================================

Execution prices may differ from displayed probabilities when liquidity is
thin or distributed across multiple book levels. We therefore report the
signed deviation of realized fills from recorded reference prices as an
execution diagnostic. In the original population analysis, PML2 exhibits a
mean adverse deviation of \(4.40\%\) of initial capital and a median of
\(1.22\%\), whereas the corresponding V3 mean is close to zero. This difference
does not establish superior execution in V3: the two backends use different
reference-price constructions, and the PML2 measure may combine spread
crossing, depth consumption, latency, and reference-price staleness.

\begin{figure}[t]
    \centering
    \includegraphics[width=\linewidth]{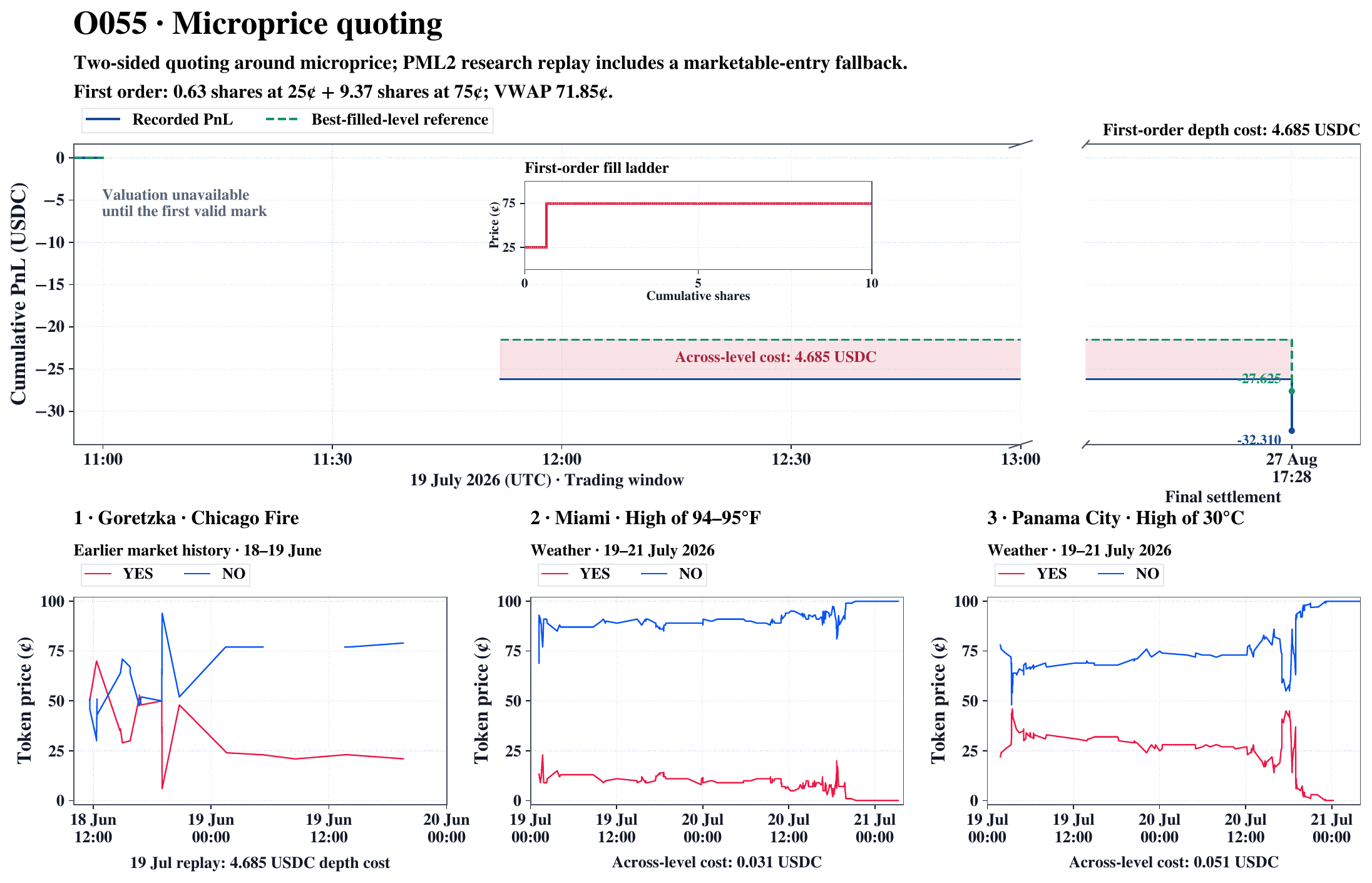}
    \caption{
    Depth-dependent execution costs for the Q055 microprice-quoting strategy.
    The upper panel shows an example fill ladder and cumulative PnL; the lower
    panels illustrate heterogeneous YES/NO price paths across sports and
    weather contracts.
    }
    \label{fig:slippage}
\end{figure}

Figure~\ref{fig:slippage} illustrates the mechanism directly. In the Q055
example, the first order fills \(0.63\) shares at \(25\) cents and another
\(9.37\) shares at \(75\) cents, producing a VWAP of \(71.85\) cents and an
across-level diagnostic cost of \(4.685\) USDC. The example shows why a quoted
probability should not be treated as the executable price of an arbitrary
trade size.

% ============================================================
\subsection{Microstructure Correlates}
\label{app:microstructure_correlates}
% ============================================================

The cross-sectional statistics exhibit two broad dependence structures.
Trading count, notional volume, and activity-cluster measures are strongly
correlated (\(\rho\approx0.91\)--\(1.00\)), suggesting that they capture a
common activity dimension. Price-response measures form a second cluster:
median price jump is strongly associated with median impact
(\(\rho=0.94\)), while the 95th-percentile jump is positively related to
large-move event rates. Effective depth is negatively associated with both
tail price jumps and event rates, consistent with larger discontinuous moves
occurring in thinner markets. These relationships are descriptive rather than
causal.

\begin{figure}[t]
    \centering
    \includegraphics[width=\linewidth]
    {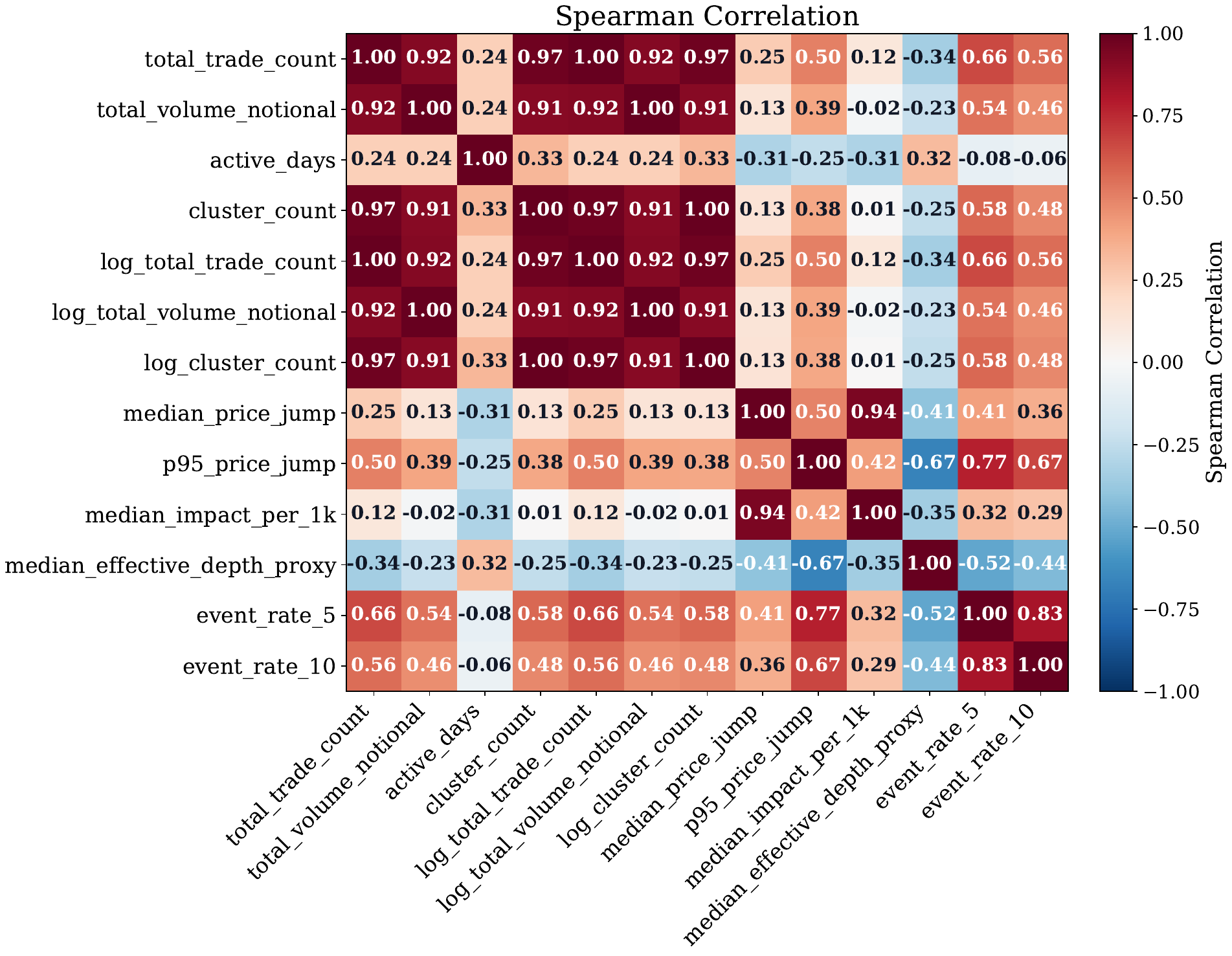}
    \caption{
    Spearman rank correlations among prediction-market activity, price-impact,
    effective-depth, and event-rate measures.
    }
    \label{fig:spearman_microstructure}
\end{figure}

% ============================================================
\subsection{Paired Fee-Aware Replay and Execution Sensitivity}
\label{app:paired_cost_replay}
% ============================================================

The preceding experiments operate primarily on frozen fills. We therefore
conduct an additional controlled replay study in which fees and execution
interventions are passed through the native account and matching logic.
The study contains \(529\) completed configurations and three deterministic
replication checks, spanning 15 strategy tasks and 19 task--market windows
(11 V3 and 8 PML2). Thirteen windows form the pre-specified core panel;
additional O072 and O055 windows are retained as explanatory cases.
Three selected windows remain data-pending and are not replaced by zero-return
observations.

Candidate code, historical inputs, lifecycle definitions, and random seeds are
held fixed within each paired comparison. Experimental interventions vary the
fee schedule, submitted order size, initial capital, entry latency, or
execution-capacity profile. The fee coefficients are controlled research
assumptions rather than historical market-specific charges, and repeated
configurations over the same window are not treated as independent market
observations.

\paragraph{Default capital: direct fee accounting is accurate.}
With \(1{,}000\) USDC initial capital and the original submitted size, the
economic execution path is unchanged across the core fee scenarios. Native
fee-aware PnL agrees with a fill-by-fill frozen-path fee calculation within
\(10^{-9}\) USDC. At this capital level, a flat 100-bps fee reduces mean V3
PnL by \(0.559\) USDC, while the PML2 nonlinear schedule with \(r=0.07\)
reduces mean PnL by \(1.513\) USDC. The two values are not directly comparable
because the task cohorts and fee functions differ.

\begin{figure}[t]
    \centering
    \includegraphics[width=\linewidth]
    {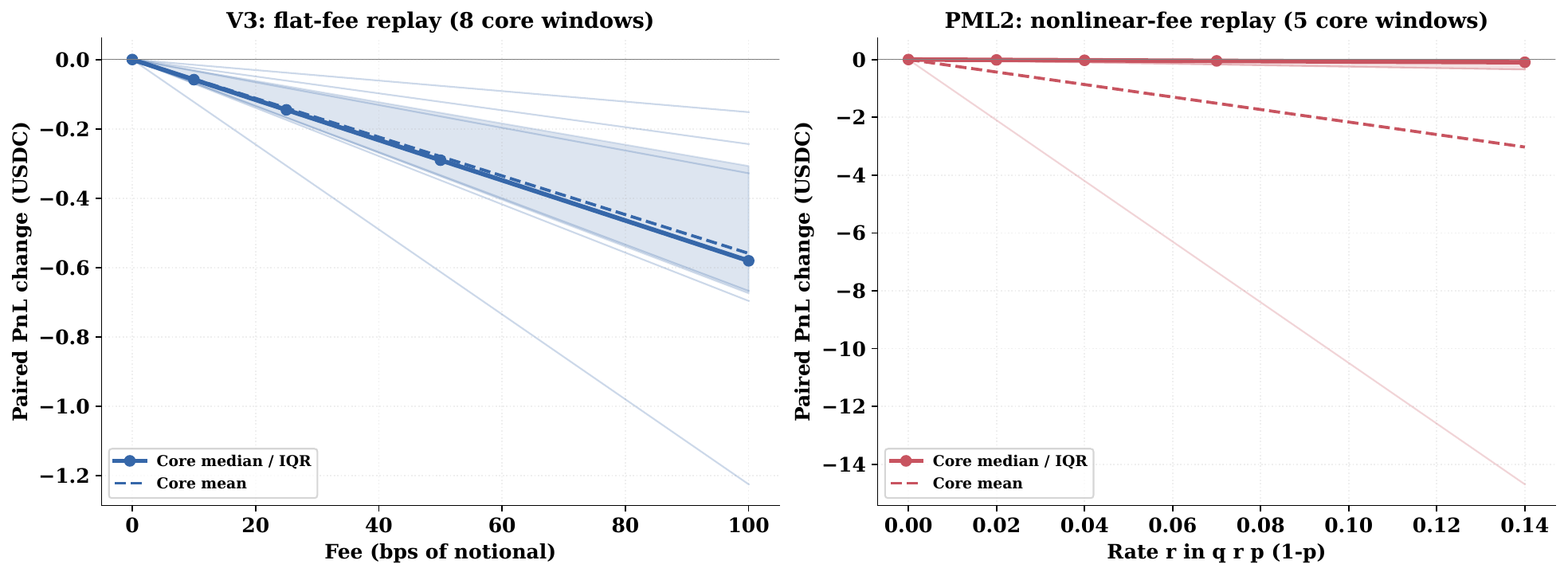}
    \caption{
    Fee-aware paired replay under default capital. Thin lines denote
    individual core windows, solid lines the median, dashed lines the mean,
    and shaded regions the interquartile range. V3 and PML2 use different fee
    parameterizations and are shown as separate sensitivity analyses.
    }
    \label{fig:fee_aware_replay}
\end{figure}

\paragraph{Capital constraints create execution feedback.}
The frozen-path approximation can fail when fee payments reduce the cash
available for subsequent trades. Eight tested fee configurations exhibit a
changed economic execution path across six task--market windows. For example,
with 20 USDC initial capital and a 100-bps fee, TASK-C087-V01 changes from a
frozen-path estimate of \(-0.20\) USDC to \(-8.99\) USDC under native replay.
Conversely, for TASK-C032-V01 with \(r=0.07\), reduced available cash prevents
some subsequent losing fills, making the native replay less negative than the
frozen-path estimate. The latter is an execution-path effect rather than
evidence that fees create profitable opportunities.

\begin{figure}[t]
    \centering
    \includegraphics[width=\linewidth]
    {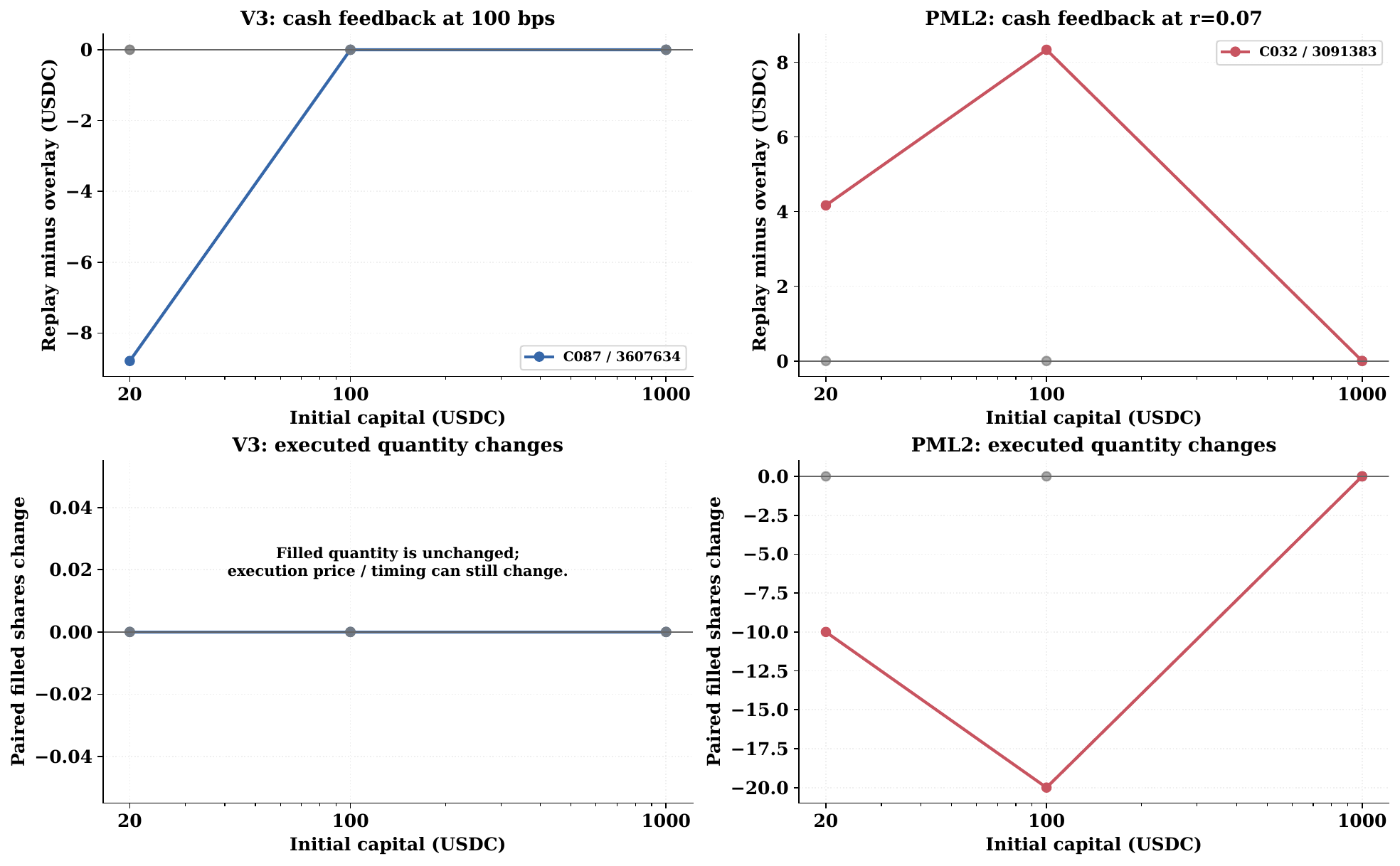}
    \caption{
    Fee-induced account feedback under capital constraints. The upper panels
    compare native replay with frozen-path accounting; the lower panels show
    changes in executed quantity.
    }
    \label{fig:cash_feedback}
\end{figure}

\begin{figure}[t]
    \centering
    \includegraphics[width=\linewidth]{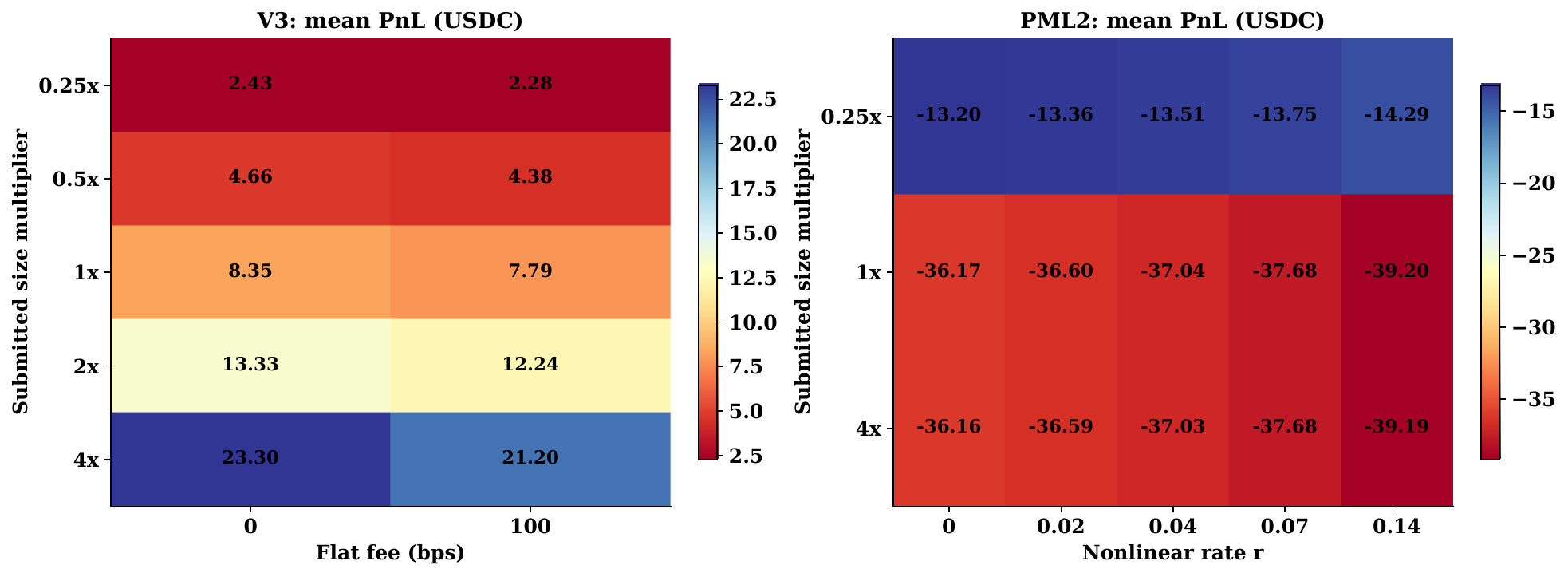}
    \caption{
    Interaction between transaction costs and submitted order size in paired
    historical replay. Left: V3 mean PnL under zero and 100-bps flat fees
    across submitted-size multipliers. Right: PML2 mean PnL under nonlinear
    fees $q r p(1-p)$ across size multipliers. Each cell reports mean account
    PnL over the corresponding tested windows. V3 and PML2 use different
    replay cohorts and fee specifications and should therefore be interpreted
    as separate sensitivity analyses rather than a cross-engine performance
    comparison.
    }
    \label{fig:fee_size_interaction}
\end{figure}

\paragraph{Fee--size interaction.}
Figure~\ref{fig:fee_size_interaction} examines how transaction costs interact
with submitted order size in the paired replay experiments. In V3, increasing
the size multiplier raises mean PnL over the tested windows, while imposing a
100-bps flat fee reduces PnL at every size; the fee penalty also grows in
absolute terms as submitted size increases. PML2 exhibits a different pattern:
mean PnL is negative throughout the tested panel and declines monotonically as
the nonlinear fee rate $r$ increases. Increasing submitted size from $0.25\times$
to $1\times$ materially changes the outcome, whereas the $1\times$ and
$4\times$ rows are nearly identical, indicating that larger requested size does
not necessarily translate into proportionally greater executed exposure when
historical book liquidity constrains execution. These results show that order
size and fees should be evaluated jointly rather than as independent
post-processing adjustments. The absolute PnL levels of V3 and PML2 are not
directly comparable because the two panels use different candidate windows,
execution models, and fee parameterizations.

\paragraph{Order size, latency, and execution capacity.}
Execution sensitivity is strategy- and market-dependent. In the O055 depth
example, increasing submitted size from \(0.25\times\) to \(1\times\) and
\(4\times\) raises the within-execution ladder-cost diagnostic from
\(0.935\) to \(4.685\) and \(19.685\) USDC, yet all three configurations
ultimately acquire the same 43.5 shares and finish with the same
\(-32.31\) USDC PnL. The ladder-cost difference therefore cannot be treated as
an additional strategy loss: its reference changes as liquidity is consumed.

Other windows do show changes in both executed quantity and PnL when submitted
size or synthetic capacity is varied. PML2 latency effects are smaller in this
sample: increasing entry delay from 200 to 3,000 ms changes one C029 window
from 125.57 to 123.50 filled shares and from \(2.511\) to \(2.470\) USDC PnL,
while the remaining tested windows show no corresponding terminal-PnL change.
Given the sparse event histories and minute-scale decisions, these results do
not constitute a calibration of millisecond live-trading latency.

\begin{figure}[t]
    \centering
    \includegraphics[width=\linewidth]
    {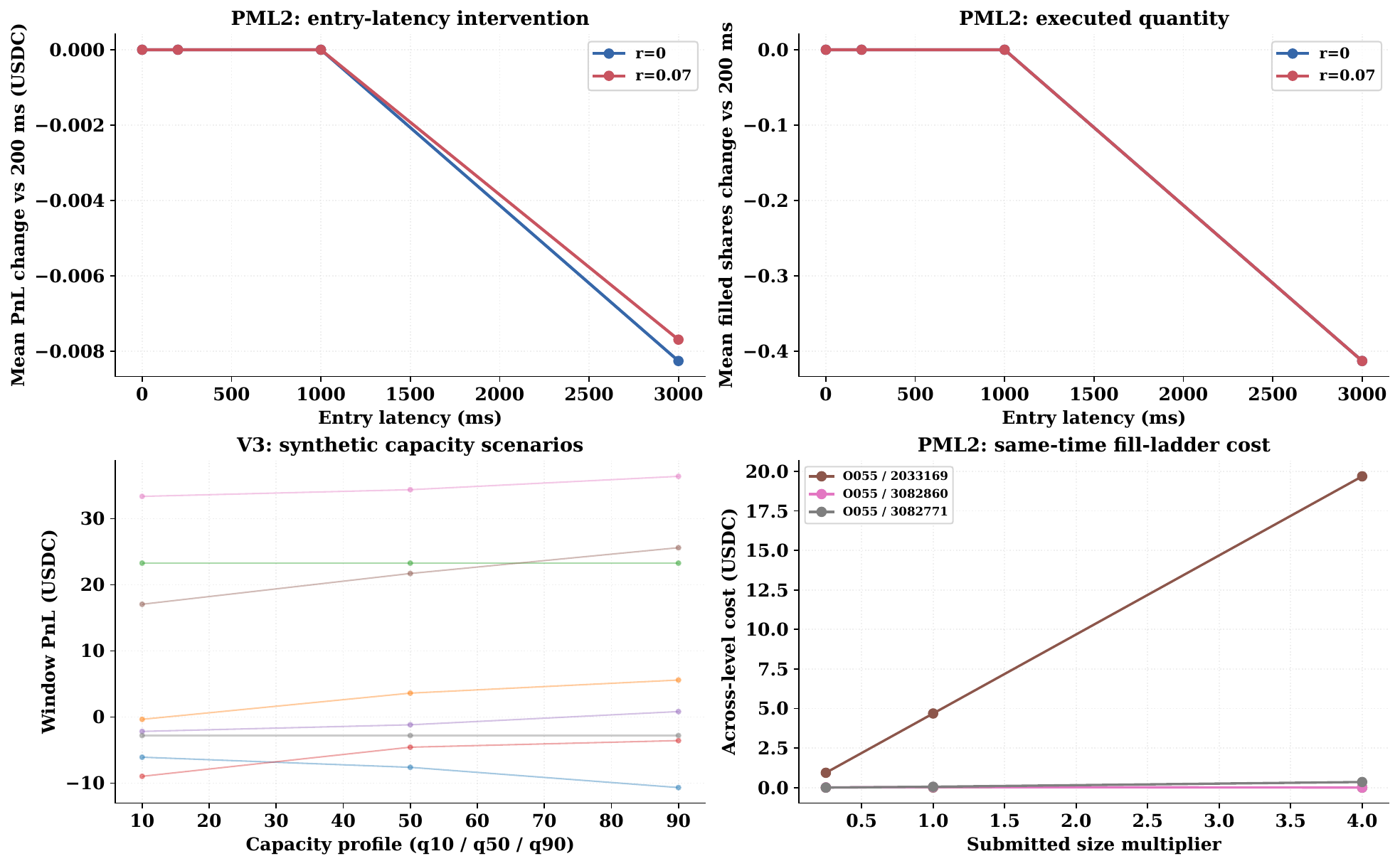}
    \caption{
    Sensitivity to entry latency, synthetic execution capacity, and submitted
    order size. Within-execution ladder cost is an execution diagnostic rather
    than an additional PnL deduction.
    }
    \label{fig:execution_sensitivity}
\end{figure}

\paragraph{Decision-time reference audit.}
The paired replay provides a stricter view of reference-price quality. Among
162 filled PML2 orders in the default zero-fee runs, the median decision-book
age is approximately \(394\) seconds. Only 30 orders have a non-crossed,
two-sided book no older than 120 seconds and therefore support a relatively
fresh decision-midpoint decomposition. Even for these observations, the
difference between fill price and decision midpoint may combine quoted spread,
book-depth consumption, residual liquidity, and repricing between decision and
execution.

Only three of the 162 filled orders exhibit a nonzero same-execution ladder
cost, totaling approximately \(4.77\) USDC, primarily from the deliberately
selected O055 explanatory case. We therefore do not interpret this quantity as
a market-wide estimate of average depth cost.

\begin{figure}[t]
    \centering
    \includegraphics[width=\linewidth]
    {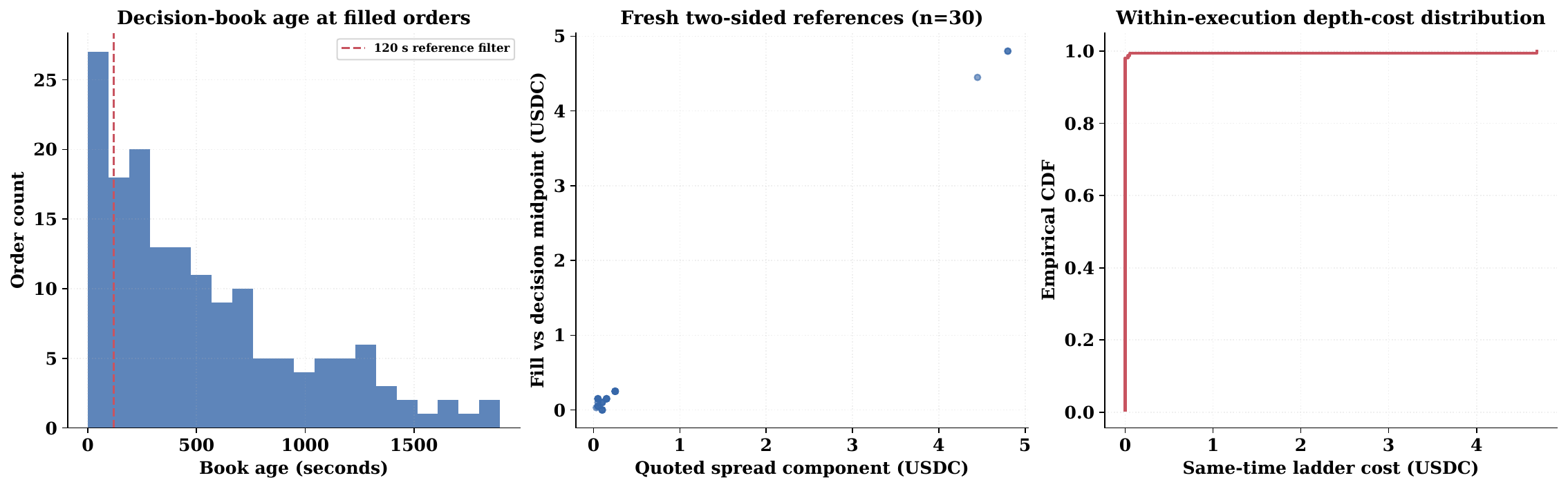}
    \caption{
    Decision-time reference and depth audit. Left: book-age distribution for
    filled PML2 orders. Center: fill deviation versus quoted-spread component
    for fresh two-sided references. Right: empirical distribution of
    same-execution ladder cost.
    }
    \label{fig:reference_depth_audit}
\end{figure}

\paragraph{Population-scale versus native-replay evidence.}
The two analyses address different questions. The population-scale sweep over
\(213{,}263\) previously evaluated episodes provides broad accounting
sensitivity under frozen fills. The 529 paired replays test whether the same
conclusion survives when fees and execution interventions alter the account
state and subsequent matching. Under ample capital, the two approaches agree
closely; under tighter capital constraints, the path itself can change.

\paragraph{Scope.}
These experiments are execution-sensitivity analyses rather than historical
fee reconstruction. They do not establish the actual fee schedule of every
Polymarket contract, validate maker rebates, calibrate live millisecond latency,
or extend the cost matrix to all 477 benchmark tasks. Likewise, the V3
capacity profiles are synthetic execution assumptions rather than empirical
LOB quantiles. The main conclusion is narrower: transaction fees and execution
frictions can affect prediction-market strategies through both direct cost
and endogenous changes in feasible execution, and the latter cannot always be
captured by an ex-post fee deduction.

\section{Computational Details}

Our experimental platform runs Ubuntu 22.04.5 LTS on a server equipped with two AMD EPYC 9554 processors, providing 128 physical CPU cores and 256 hardware threads, together with approximately 1.1 TiB of system memory. Agent-based strategy design, semantic review, and code generation use a locally deployed Qwen3.8-27B model \citep{qwen2026qwen38} served through SGLang. The inference service uses two NVIDIA GeForce RTX 5090 GPUs, each with 32 GB of memory, with tensor parallelism set to two. The deployment supports a maximum context length of 32,768 tokens and processes one active inference request at a time. Requests disable thinking mode and require structured JSON responses.

\section{ETHICAL STATEMENT}
No human subjects were involved in this study. All employed models are open-source and used in compliance with their licenses for research purposes only. The proposed method entails no significant privacy or security concerns.

\section{Use of Large Language Models}
After the initial draft is completed, we will use specific words to have the LLM refine the text.

\end{document}